\documentclass[english,
notitlepage,
amsmath,
amssymb
]{revtex4-1}

\usepackage{graphicx}
\usepackage{dcolumn}
\usepackage{bm}
\usepackage{xcolor}         

\DeclareMathOperator{\Tr}{Tr}

\begin{document}


\title{Machine learning at the mesoscale: a computation-dissipation bottleneck}

\author{Alessandro Ingrosso}
\author{Emanuele Panizon}%
\affiliation{
Quantitative Life Sciences, Abdus Salam International Centre for Theoretical Physics, 34151 Trieste, Italy}%

\date{\today}

\begin{abstract}
The cost of information processing in physical systems calls for a trade-off between performance and energetic expenditure. Here we formulate and study a computation-dissipation bottleneck in mesoscopic systems used as input-output devices. Using both real datasets and synthetic tasks, we show how non-equilibrium leads to enhanced performance. Our framework sheds light on a crucial compromise between information compression, input-output computation and dynamic irreversibility induced by non-reciprocal interactions.
\end{abstract}


\maketitle

What does a theory of computation at the mesoscopic scale look like?
To begin to answer this question, we need to bridge the formalism of computation with a physical theory of systems whose energy scales are close to thermal fluctuations. Stochastic Thermodynamics (ST), by associating single stochastic trajectories with meaningful thermodynamic quantities~\cite{seifert_entropy_2005,seifert_stochastic_2012,van_den_broeck_ensemble_2015,peliti_stochastic_2021}, exposes the deep relation between information and dissipation. One of the fundamental results of ST is that information and time irreversibility, as measured by the rate of Entropy Production (EP)~\cite{andrieux2007entropy, parrondo2009entropy}, are inherently related~\cite{Landauer_irreversibility, bennet_notes,esposito_second_2011,berut_experimental_2012,parrondo_thermodynamics_2015}. Thermodynamic Uncertainty Relations~\cite{barato_thermodynamic_2015,seifert_stochastic_2018,koyuk_thermodynamic_2020,horowitz2020thermodynamic}
have been derived that describe fundamental precision-dissipation trade-offs, leading to a framework successfully applied to a variety of bio-chemical processes, such as chemo-sensing~\cite{lan2012energy,sartori_thermodynamic_2014,ngampruetikorn_energy_2020}, copying~\cite{sartori_thermodynamics_2015}, reaction networks~\cite{barato_thermodynamic_2015,espositoReaction2016,fritz_stochastic_2020}, cascade models of synapses~\cite{karbowski_metabolic_2019}, among others.

To set the following discussion, we will refer to computation at the mesoscopic scale as the ability of a system to react to the environment - via physical interactions between its parts and external heat baths - in such a way that the modification of its state depends on some function of the environmental conditions. This transformation possibly leads the system far from equilibrium.

Encoding external signals in their entirety is one of such computations. Borrowing terminology from Machine Learning (ML), a mesoscopic system can be considered as an ``autoencoder'', thus focusing on its ability to sense, compress information and perform error correction capabilities~\cite{barato2013information, barato2014efficiency}.

Full encoding, however, may be energetically wasteful when a computation regards a limited aspect of the environment:  discarding non-relevant information allows to strike a balance between performance and energy expenditure, in a manner crucially dependent on the task at hand. We recognize this task-dependence of the performance/cost trade-offs as the critical ingredient of any physical theory of computation.

On one side of such trade-off lies dissipation, the study of which is starting to be addressed in many body systems~\cite{herpich_collective_2018,sune_out--equilibrium_2019,koyuk_thermodynamic_2020,herpich_stochastic_2020-1}. Irreversibility of macroscopic neural dynamics is also attracting attention~\cite{cofre_information_2018,cofre_introduction_2019,lynn_emergence_2022,lynn_decomposing_2022}.

The system's computational performance, on the opposite side of the trade-off, can be formulated both in information theoretic terms and in the more practical lens of standard error metrics employed in ML.
One recently emerging approach attempts to define a framework for irreversibility in formal models of computational systems~\cite{wolpert_spacetime_2019,wolpert_stochastic_2019,wolpert_thermodynamics_2020}, in a way that is agnostic to physical implementations.

Here, we consider generic parametrizations of mesoscopic systems whose stochastic transitions are induced by an environment, possibly out-of-equilibrium, so that resulting interactions may show non-reciprocity~\cite{ivlev2015statistical}. In particular, we focus on asymmetric spin models, which have been subject of intense study in the field of disordered systems~\cite{crisanti_dynamics_1987,crisanti_dynamics_1988,aguilera_unifying_2021,aguilera_nonequilibrium_2022} and provide a bridge to classical models of neural computation~\cite{ginzburg_theory_1994,renart_asynchronous_2010,roudi_multi-neuronal_2015,dunn_correlations_2015,shi_spatial_2022}.

In line with conventional formalism of neural networks, we consider the dynamics of these systems as producing internal representations of their inputs, the geometry and intrinsic dimensionality of which impact the ability to learn input-output relations. We show how entropy-producing non-reciprocal interactions ~\cite{Loos_2020,Loos2023} are crucial to generate effective representations, in such a way that a fundamental trade-off emerges between expressivity and performance.

\section{\label{sec:bottleneck}A computation-dissipation bottleneck}
The stochastic dynamics of mesoscopic systems, usually described using continuous-time Markov processes, results from their interactions with thermal baths and external driving mechanisms. Let us consider a system $\mathcal{S}$ with discrete states $s$, driven by a time homogeneous input protocol $x$. The evolution of the probability of state $p\left(s,t\right)$ is given by a master equation with jump rates $k_{s's}$ from state $s$ to states $s'$.
To facilitate the connection to ML, we take a set of parameters $\theta$ that determine -- rather abstractly -- the jump rates.

We assume computation is performed on a timescale much longer than any initial transient. For each independent input $x$, the system reaches a steady-state (SS) probability $p(s|x)$, serving as internal representation of $x$. At the (possibly non-equilibrium) SS, each input $x$ is associated to an average EP rate, $\Sigma\left(x\right)$, a measure of irreversibility at the steady state and corresponding to the housekeeping heat.
In Markovian systems with discrete states, the EP rate can be computed via the Schnakenberg formula~\cite{Schnakenberg_formula,RoldanEstimating2010}: 
\begin{equation}
   \sigma = \frac{1}{2}\sum_{s,s'} J_{ss'} \log\frac{k_{ss'} p\left(s' \right)}{k_{s's} p\left(s \right)}
\end{equation}
where $J_{ss'} = \left[ k_{ss'} \, p\left(s'\right) - k_{s's} \, p\left(s\right)\right]$ are the steady state fluxes and we work in units where the Boltzmann constant $\kappa_{B}=1$.
Note that in our case $\sigma = \sigma(x, \theta)$
through $k_{ss'}, J_{ss'}$ and $p(s)$.

A supervised learning task is specified by a finite set $\mathcal{D}=\left(x,y\right)$ of input-output pairs, so the EP rate averaged over the whole dataset is simply $\Sigma (\theta) = \frac{1}{\left|\mathcal{D}\right|} \sum_x \sigma \left( x , \theta\right)$. Alternatively, the learning task could be defined by a distribution over the input space $p\left(x\right)$ and an conditional output distribution $p \left(y|x\right)$. The (average) EP rate is similarly $\Sigma(\theta) = \sum_x p\left(x\right) \sigma\left( x , \theta\right)$.

The EP rate is a function of the dynamic process alone. How the resulting $p(s|x)$ is able to disentangle and predict the output is a separate, task-specific factor.

There exist a number of ways to define a good measure of computational performance. One possible choice is the mutual information between the internal representations $s$ and the output $y$, i.e. $I(s,y)$: such choice makes no assumption on the additional computational burden needed to extract the information about $y$, possibly encapsulated in arbitrarily complex high-order statistics of the steady state distribution. A different path is to use a small subset of moments of $p(s|x)$ as internal representations of the inputs, to be then fed to a simpler linear readout. This approach, closer to standard ML practice, allows us to use the Mean Square Error (MSE) or Cross Entropy (CE) loss functions. Both approaches and their limitations will be explored in the following.

Given a performance measure $\mathcal{G}(\theta)$, the trade-off can be encapsulated in a quantity:
\begin{equation}
    \mathcal{L}\left( \theta \right) = \mathcal{G}(\theta) - \alpha \, \Sigma\left(\theta\right),
    \label{eq:cdb}
\end{equation}
where $\alpha$ is a positive parameter that has units of time. We study the trade-off by optimizing $\mathcal{L}$ over the interaction parameters $\theta$ for different values of $\alpha$: increasing $\alpha$, the cost of dissipation with respect to performance is enhanced, with the $\alpha\to\infty$ limit effectively constraining the system to be at equilibrium.

In this letter, we first use numerical methods to build a multi-spin system that performs two different classification tasks. We then employ an analytically solvable 2-spin model to investigate the enhanced expressivity of non-equilibrium systems with respect to equilibrium ones, and relate it to the structure of their computational tasks.
\begin{figure}
    \centering
    \includegraphics[width=0.5\linewidth]{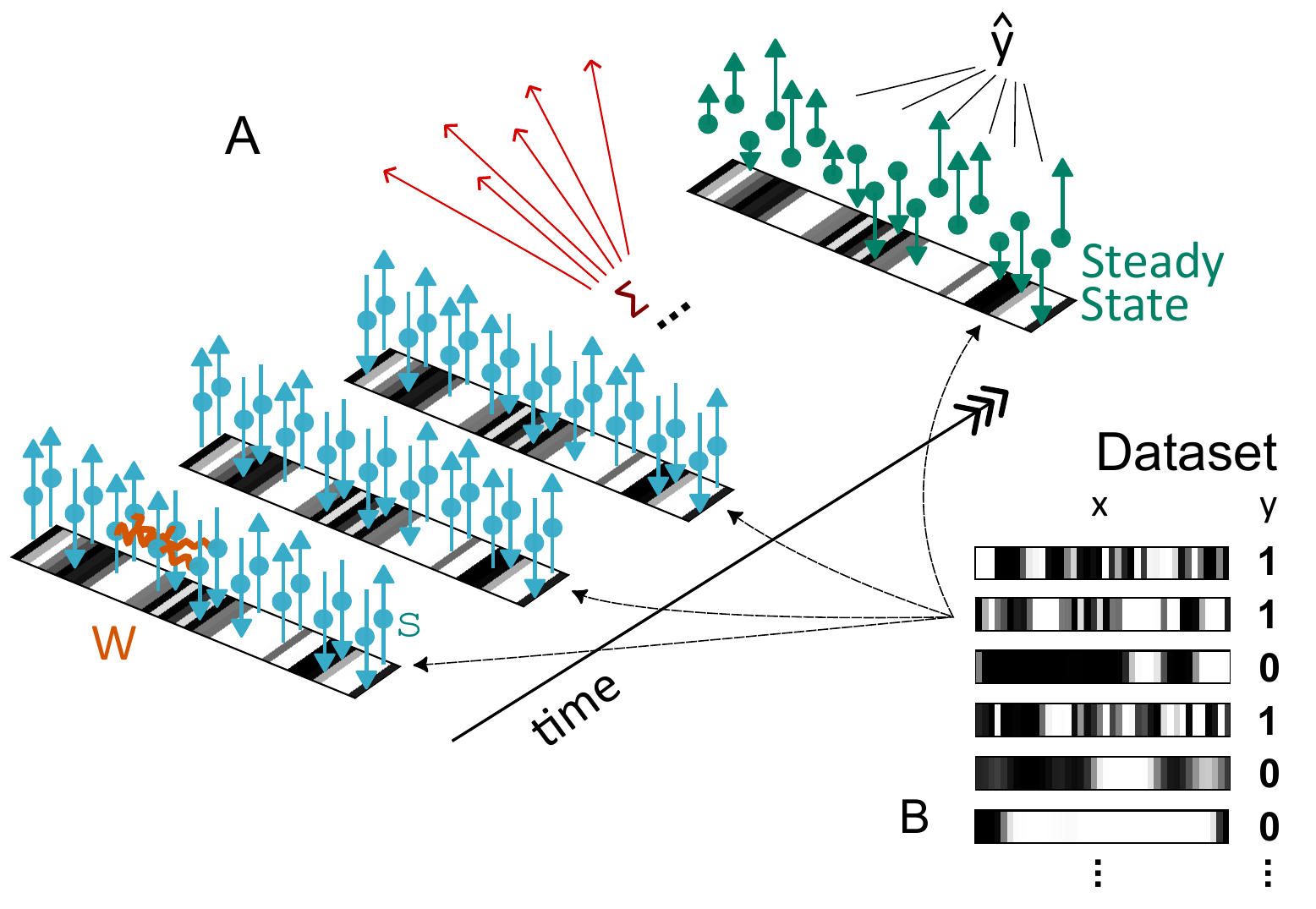}
    \caption{Schematic of a multi-spin system processing its inputs in a classification task.
    \textbf{A}: The system $\mathcal{S}$ evolves in time in the presence of constant inputs (external fields) $x_i$ and interactions couplings $W$, until a non-equilibrium steady state (NESS) $p\left(s|x\right)$ is reached. Time evolution is associated with an entropy production rate $\Sigma$. Information about the output label is then extracted from $p\left(s|x\right)$. In the simplest case, a linear readout $W_{out}$ can be used on the averages $m_x$.
    \textbf{B}: A subset of an input-output dataset $\mathcal{D}$.
    }
    \label{Fig:schematic}
\end{figure} 

\section{\label{sec:big_systems} Multi-spin systems as stochastic recurrent networks}
To exemplify the computation-dissipation trade-off, we use a spin-based model to perform an input-output computation. Specifically, a classification task where inputs $x$ -- schematically represented by the tape in Fig~\ref{Fig:schematic} -- must be correctly associated with given output labels $y$. 

The system at hand is composed of two chains of size $N$ with possibly asymmetric couplings. Spins of the two chains are driven by the same inputs $x_i$, serving as constant external fields. Each spin $s_i$ is subject to random flips with rates $k^{\left(i\right)}_{s} \propto e^{-\beta s_i \left(W s + x\right)_i}$. Interactions, encoded in the matrix $W$, connect spins both along the same chain and across the two lines of spins, similarly to an implicit, stochastic version of a convolutional layer (see Appendix). When $W$ is symmetric, the system relaxes to the equilibrium of a Hamiltonian $\mathcal{H}=-\frac{1}{2}s^T W s - x$ at inverse temperature $\beta$. Non-reciprocal interactions ($W \neq W^T$) lead to non-equilibrium and a non-zero EP rate.
After a transient, the system reaches a steady state $p\left(s|x\right)$, with an average magnetization $m_x = \left< s |x \right>$ and an entropy production rate $\sigma\left(x, \theta = W\right)$. For any input-output dataset, each $W$ will thus be associated with both a different task performance and an average EP rate $\Sigma$.

In close analogy with standard ML methods, we implement a final linear readout of the average magnetization $ W_{out} m_x$, with a learnable matrix $ W_{out}$. This allows us to separately consider the system's computation as a two-step process: (i) a highly non-linear deformation of the input space $x$ into $m_x$ induced by the dynamics of the process, akin to what occurs in a hidden layer of an artificial neural network; (ii) a separation in the $m_x$ space to produce the output $y$. Note that our formalism is a stochastic, mesoscopic generalization of the recently introduced implicit layers, which serve as building blocks of deep equilibrium odels~\cite{Bai_deep,Bai_multiscale}.

\begin{figure}
    \centering
    \includegraphics[width=0.5\linewidth]{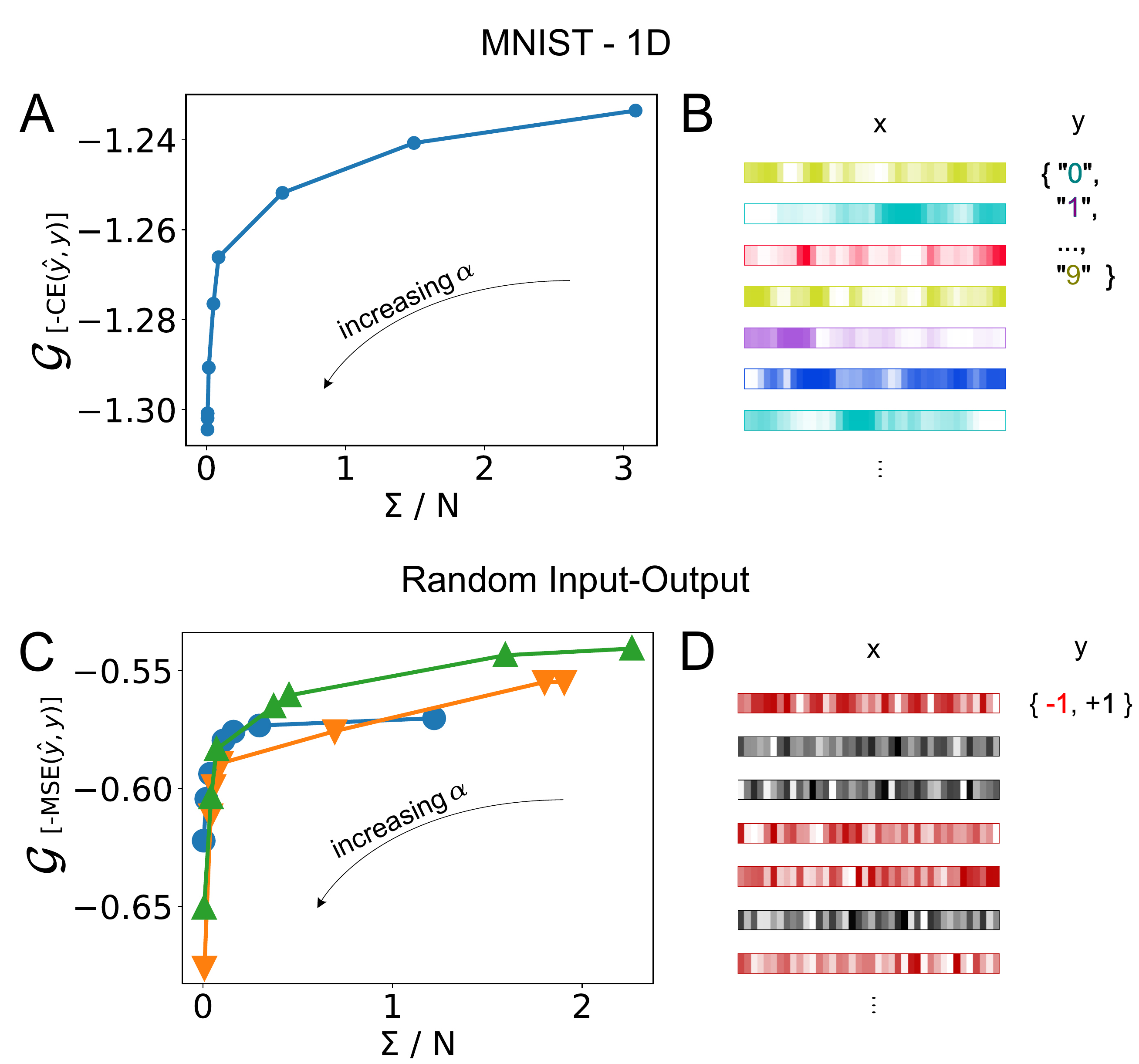}
    \caption{Computation-dissipation bottleneck for a network solving a classification task at steady state.
    \textbf{A}: Cross Entropy Error (CE) vs normalized Entropy Production rate $\Sigma/N$ on the MNIST-1D dataset ($M=4000$ datapoints, $10$ labels, $N=40$) of a system composed of two spin-chains with interactions up to the 2nd nearest neighbor.
    \textbf{B}: Schematic of the MNIST-1D task.
    \textbf{C}: Mean Square Error (MSE) vs Entropy Production Rate $\Sigma/N$ normalized by the number of spins on the Random input-output task with $M=100$ input patterns and $2$ labels. Each curve is the minimum over $10$ independent initialization seeds in a Gillespie-based optimization algorithm. $10$ different realizations of the task are shown.
    \textbf{D}: Example realizations of the Random dataset.
    }
    \label{Fig:multi_spins}
\end{figure}

To minimize $\mathcal{L}$, we coupled a standard Gillespie algorithm~\cite{gillespie_stochastic} for the simulation of the system's evolution with each input field $x$ to a gradient-based optimization method. Due to the stochastic nature of the Gillespie trajectory and the high dimension of the $W$ parameter space, we adopted a finite-difference method called Simultaneous Perturbation Stochastic Approximation (SPSA)~\cite{spall_overview_1998} to compute an estimate of the gradient (see Appendix for details). The solutions at each value of $\alpha$ allow us to construct an optimal front between $G^*(\Sigma)$ and $\Sigma^*(\alpha)$, where $^*$ denotes that optimal values of Eq.~\ref{eq:cdb}, as shown in Fig.~\ref{Fig:multi_spins}A,C.

We showcase our approach with two different tasks. The first is MNIST-1D~\cite{greydanus2020scaling}, a one-dimensional version of the classic digit-classification dataset MNIST. Each element has an input dimension $N=40$ and belongs to one of $10$ different classes, i.e. the digits. See an example of the input configurations in Fig.~\ref{Fig:multi_spins}B. To enable multi-label classification, we apply a normalized exponential function $SM$ (softmax) to the output to get a $10$-dimensional probability vector $\hat{y} = SM(W_{out} m_x)$, and use the negative cross-entropy between actual labels and $\hat{y}$, $\mathcal{G} = -CE\left( \hat{y}, y \right)$, as a measure of task performance.

Our results show an inverse relationship between task performance and entropy production at steady state, Fig.~\ref{Fig:multi_spins}A. Enforcing the system to be at equilibrium ($\alpha \rightarrow \infty$) reduces performance by $\approx 5\%$ and accuracy -- defined as the percentage of output labels identified as most probable -- by $7\%$. This highlights how non-reciprocal interactions enhance the complexity of internal representations needed for learning, at the cost of higher dissipation.

The second task is a classic random input-output association~\cite{Gardner_optimal,Gardner_unfinished,engel_van_den_broeck_2001}, where input components $x_i^\mu$ of each pattern $\mu=1,...,M$ are drawn i.i.d. from a normal distribution, and labels are random $y\in\left\{ -1,+1\right\}$ with probability $1/2$ (Fig.~\ref{Fig:multi_spins}D). We measure the performance in this task by the mean squared error (MSE): $\mathcal{G} = -MSE\left(\hat{y}, y \right)$, where $\hat{y} = W_{out} m_x$.
For all random instances of this second task, we reproduce the front between entropy production and performance, Fig.~\ref{Fig:multi_spins}C. While quantitative details differ slightly for different instances, the performance consistently increases with the amount of non-reciprocity in the optimal coupling matrix $W$ and therefore with dissipation in the system.

\section{A tractable 2-spin system}
\begin{figure}
    \centering
    \includegraphics[width=0.5\linewidth]{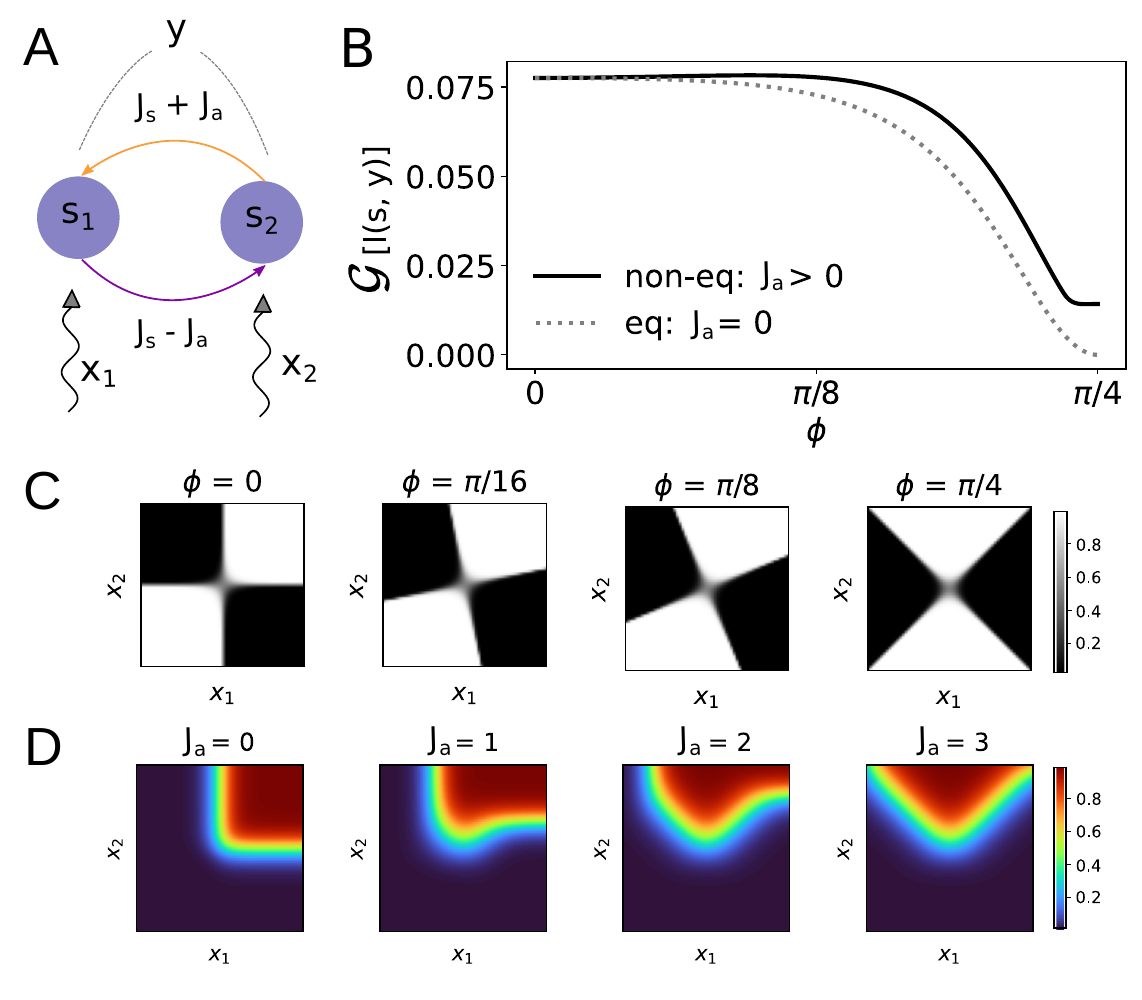}
    \caption{
    \textbf{A}: Schematic of a 2-spin system driven by input external fields $x_1$ and $x_2$.
    \textbf{B}: Conditional probability $p\left(y = 1 | x\right)$ for the family of input-output tasks described in the main text for different values of the rotation parameter $\phi$ and $\eta=2$. Inputs $\left\{ x_{1},x_{2}\right\} $ are extracted from two independent Normal distributions.
    \textbf{C}: Mutual information between input $x$ and output $y$ as a function of rotation angle $\phi$. The optimal solutions for equilibrium (non-equilibrium) systems are shown with a dashed grey (continuous black) line. Parameters: $\beta=3$, $\eta=3$.
    \textbf{D}: Steady-state probability for the state $s=(+1,+1)$ for $J_s=0$ and increasing values of the non-reciprocity strength $J_a$. The range is $[-2.5, 2.5]$ for both inputs $x_1$ and $x_2$.    
    }
    \label{fig:task-dependence}
\end{figure}

To exemplify a general formulation of the computation-EP bottleneck, let us study a specific case of the system we introduced in the previous section, which can be solved analytically. We consider a 2-spin system with asymmetric couplings $\theta = (J_s + J_a, J_s - J_a)$, driven by constant two-dimensional inputs $x$ that act as external fields, Fig.~\ref{fig:task-dependence}A. When $J_a=0$, the system respects detailed balance and reaches an equilibrium state. Non-reciprocity in the coupling between the spins leads to non-negative $\Sigma$.

The information-coding capabilities at steady state of such a system have been recently analyzed~\cite{ngampruetikorn_energy_2020}. In turn, we treat such a mesoscopic network as an input-output device. 
In full generality, we prescribe a stochastic rule by a known conditional distribution $p\left(y | x \right)$, with $y\in\{0,1\}$ a binary output variable. This formulation encompasses classic Teacher-Student setup~\cite{schwarze1992generalization,stat_mech_seung_sompo,engel2001statistical,baldassi_subdominant_2015} and mixture models~\cite{Louriero_mixture,refinetti2021classifying} employed in the theoretical study of feed-forward neural networks.
At variance with the previous examples, we relax the assumption of a linear readout and ask how much information $I\left(s,y\right)$ about the output $y$ is contained in the steady state probabilities $p\left(s | x\right)$. 

Let us consider a stochastic and continuous generalization of a parity gate, where the output is prescribed by $p\left(y=1|x\right)=\text{sigmoid}\left(\eta x^{\phi}_{1} x^{\phi}_{2}\right)$, with $x^{\phi}=R^{\phi} x$, $R^{\phi}$ a rotation operator of angle $\phi$. This defines a family of tasks with a controllable degree of asymmetry in input space. Examples of such tasks are shown in Fig.~\ref{fig:task-dependence}C. The additional parameter $\eta$ affects the sharpness in the change of the output probability as a function of $x$. 

The mutual information $I\left(s, y\right ) = H\left(y\right ) - H\left(s|y\right)$ can be computed easily using the conditional independence of $y$ and $s$ given $x$. For $\phi=0$, the optimal structure is an equilibrium system ($J_a^\star=0$). As $\phi$ increases, the optimal 2-spin network has asymmetric weights ($J_a^\star>0$), implying a non-zero entropy production at steady state, Fig.~\ref{fig:task-dependence}B. Limiting the system to be at equilibrium thus results in performance degradation, down to a minimum of zero information when the rotation reaches $\phi = \pi / 4$.

For a given value of $\phi$ and the free parameter $\alpha$, one can define the computation-dissipation trade-off in the form of maximizing Eq.~\ref{eq:cdb}, now with $\mathcal{G} = I\left(s,y\right)$.
Note the analogy with the formulation of task-relevance of internal representations provided by the classic Information Bottleneck~\cite{tishby_information_2000,strouse_deterministic_2017,chalk_relevant_2016}. Here, instead of a compromise between input compression and retention of output information, we trade off the latter with dissipation.

\begin{figure}
    \centering
    \includegraphics[width=0.5\linewidth]{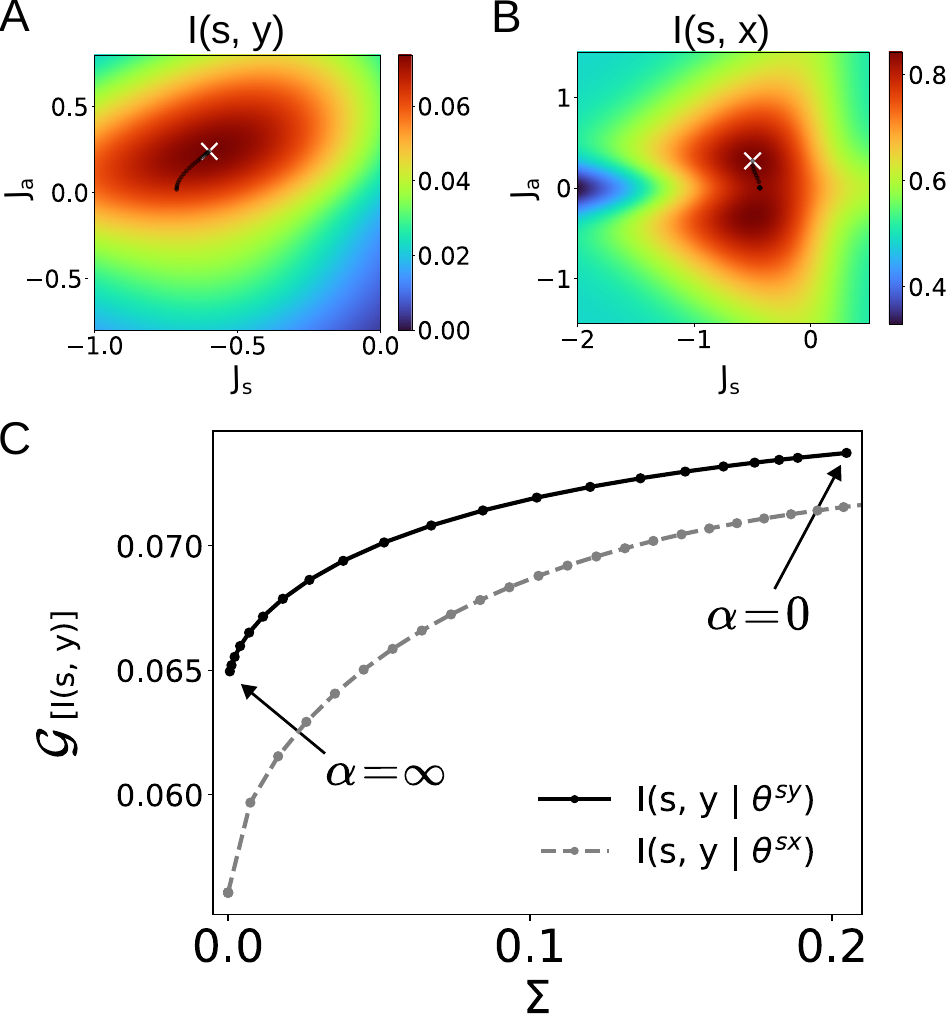}
    \caption{
     \textbf{A}: Color plot of mutual information $I\left(x,s\right)$ in the $\left\{J_s,J_a\right\}$-plane. The optimal parameter set $\theta^{sx}$ is shown for different values of $\alpha$ (white: $\alpha=0$, black: $\alpha>0$).
     \textbf{B}: Same as \textbf{B} for $I\left(s,y\right)$ and the optimal parameter set $\theta^{sy}$.
     \textbf{C}: Mutual information between input $x$ and output $y$ for $\phi = 0.5$ as a function of the entropy production rate at steady state $\Sigma$ for both $\theta^{sy}$ (black) and $\theta^{sx}$ (grey). Inputs $\left\{x_1,x_2\right\}$ are Gaussian with correlations $\rho=0.95$. Additional parameters: $\beta=3$, $\eta=3$.
     }
    \label{fig:2_spins}
\end{figure}

We can compare the performance of an auto-encoding system, whose couplings $\theta^{sx}=\left\{J^{sx}_s,J^{sx}_a\right\}$ are chosen using $\mathcal{G} = I(s,x \,|\, \theta)$, with that of a system with parameters $\theta^{sy}$ optimizing $\mathcal{G} = I(s,y \,|\, \theta)$, the information about $y$. The optima corresponding to $\alpha = 0$ have finite non-reciprocal terms $J_a$ -- see Fig~\ref{fig:2_spins}A,B -- and therefore positive, but finite, EPs. For all values of $\phi$ there exists a maximum dissipation rate above which performance degrades~\cite{baiesi2018life}.

Fig~\ref{fig:2_spins}C shows the computation-dissipation front for a task with $\phi=0.5$, each point representing a different optimal compromise between input-output performance, measured by the mutual information $I\left(s,y\right)$, and rate of entropy production at steady state. We chose a parameter regime where a non-equilibrium solution is optimal also for $I(s,x)$~\cite{ngampruetikorn_energy_2020}.
Crucially, a system that maximizes the information on the entire input $I\left(s,x\right)$ performs worse than a system tailored to maximize the output information. This is a hallmark of optimization of task-relevant information.

This simple system allows us to explore the relation between the non-equilibrium steady state probability $p(s|x)$ and the task. Fixing $J_s=0$, the effect of increasing $J_a$ resembles a rotation by an angle of $\pi/4$ of $p(s|x)$ in the region where $|x| < J_a$, see Fig.~\ref{fig:task-dependence}D.
An increasing amount of non-reciprocity in the system will thus align the steady-state probabilities $p(s|x)$ with the rotation induced on the conditional output $p\left(y | x \right)$ by the angle parameter $\phi$.

\section{\label{sec:discussion} Discussion}
We introduced a framework to characterize a fundamental trade-off between computational capabilities and energetic expenditure in mesoscopic systems. We showcase how such systems can be used in supervised learning tasks and how limiting entropy production can degrade their performance, as measured either using standard loss functions in ML or with information theoretical methods.

Our results point to the general necessity to gauge encoding and task-relevance while considering energetic trade-offs. In a simple 2-spin system, we show how non-reciprocal interactions affect the capability of the system to solve different tasks optimally, independently from the encoding of the input signals: a simple modulation of the input-output task switches the optimal system configuration from an equilibrium to a highly non-equilibrium one.

Linear stochastic systems (Ornstein-Uhlenbeck processes) are another case for which one can derive an analytical expression for the computation-dissipation trade-off (see Appendix). The emerging trade-off between entropy production and output information is again controlled by the degree of asymmetry of the task in input space.

In this study, we concentrated on one-time statistics of the steady state distribution, leaving aside interesting properties of time-correlations. The study of both transient behavior and non-stationary protocols -- where more general tasks can be formulated for instance by prescribing time-dependent average responses $y\left(t\right)$ to multi-dimensional time-dependent signals $x\left(t\right)$ -- opens an interesting avenue to investigate general speed-dissipation-computation trade-offs within this framework. Special care must be used
in such cases to distinguish between housekeeping and excess entropy production~\cite{hatano_steady-state_2001}.

Studying the impact of hidden units is an important avenue for future work. In generative models, marginalization over hidden states is the crucial ingredient to induce higher-order interactions. This forms the basis for the attention mechanism in transformers~\cite{vaswani2017attention} -- arguably the most powerful ML models to date~\cite{devlin2018bert, openai2023gpt4} -- as the recent works on modern Hopfield networks~\cite{hopfield_neural_1982,krotov_dense,krotov2021large,ramsauer2021hopfield,attention_allyouneed} have shown.

Drawing a bridge between ML, theoretical neuroscience and ST can prove fruitful in systematically studying how internal representations depend on the cost.
Rate-distortion approaches have been used to study the impact of information compression on classification accuracy and maximal attainable rewards~\cite{russo_information-theoretic_nodate,xu_information-theoretic_2017,russo_satisficing_2020,pacelli_task-driven_2019,pacelli_learning_2020,majumdar_fundamental_2022}, but a general theory is currently lacking.
Our perspective is complementary: energetic costs are expected to have a strong impact on the complexity of internal representations, leading to different mechanisms for information processing.

\section*{Acknowledgements}
We wish to thank Antonio Celani, Roman Belousov and Edgar Roldan for fruitful discussions and for reading a preliminary version of this manuscript.

\appendix
\renewcommand\thefigure{\thesection.\arabic{figure}}    
\section*{Appendix}
\setcounter{figure}{0}

\section{Steady State and Mutual Information in a Continuous Time Markov Chain}
The evolution of the probability $p\left(s,t\right)$ of state $s$ is described by a master equation:
\begin{equation}
\frac{d}{dt}p\left(s,t\right)=\sum_{s'}\left[k_{ss'}\left(t\right)p\left(s',t\right)-k_{s's}\left(t\right)p\left(s,t\right)\right],
    \label{eq_sm:ctmc}
\end{equation}
with $k_{ss'}\left(t\right)$ the jump rate from state $s'$ to state $s$, whose time dependence is due to a generic external protocol $x\left(t\right)$. In our case with a constant-in-time protocol, the steady state $p\left(s|x \right)$ can be obtained extracting the kernel of the matrix $R_{ss'}=k_{ss'}-\delta_{s,s'}\sum_{s''} k_{s''s}$. For systems of small size, this is viable numerically using Singular Value Decomposition (SVD).

The mutual information between the input $x$ and the system state $s$ at steady state can be easily computed using $I \left(s,x\right) = H\left(x\right) - H\left(s|x\right)$, with $H$ the Shannon entropy. As for $I \left(s,y\right) = H\left(s\right) - H\left(s|y\right)$, the entropy term $H\left(s|y\right)$ can be easily obtained by exploiting the conditional independence between $y$ and $s$, which implies that the joint distribution $p\left(s,y\right)$ can be written as:
\begin{equation}
p\left(s,y\right)=\sum_{x,s,y} p\left(s,y|x\right)p\left(x\right)=\sum_{x,s,y} p\left(s|x\right)p\left(y|x\right)p\left(x\right).
\label{eq_sm:p_s_y}
\end{equation}
Using Eq.~\ref{eq_sm:p_s_y}, the posterior distribution $p\left(s|y\right)$ is directly calculated using the Bayes theorem.

\section{Training of a multi-spin systems}\label{SPSA}

\subsection{Details on the system}
We consider a system composed of two chains of size $N$. Interactions connect spins up to the $k^{th}$ neighbours, where we use $k=2$.  If we identify a spin by $\left(m,n\right)$ where $1 \leq m \leq N$ is the position in the chain and $n=1,2$ the chain index, two spins $\left(m_i,n_i\right)$ and $\left(m_j,n_j\right)$ are connected if $|m_i-m_j| \le k$. The interaction parameter $W_{ij}$ depends only on $m_i-m_j$ and $n_i-n_j$, so that the number of non-zero, fully independent parameters of $W$ is $8k-2$. The external input $x$ is repeated such that it is the same for both chains.
Such spin system at steady state implements a stochastic version of an implicit convolutional layer with two channels~\cite{Bai_deep,Bai_multiscale}.

\subsection{Datasets}
MNIST-1D is a 1-dimensional version of size $N=40$ of the classic MNIST handwritten digits dataset~\cite{greydanus2020scaling}. We used 4000 training samples, organized in $10$ different classes, each containing roughly $400$ samples.
Data is available at https://github.com/greydanus/mnist1d, where a description of its generation from the original MNSIT dataset is given.

We generated instances of the Random Task by drawing $M=100$ patterns $x^\mu$ in dimension $N=10$, with components $x_i^\mu$ independently from a Normal distribution. The corresponding labels $y^{\mu}$, drawn from $\left\{ -1,+1\right\}$ with probability $1/2$, were randomly associated with each pattern.

\subsection{Details on Gillespie simulations}
The Gillespie algorithm~\cite{gillespie_general,gillespie_stochastic} offers a remarkably simple method to generate stochastic trajectories by randomly selecting sequences of jumps between states. Let us consider a system with a discrete number of states $s$ and transition rates $k_{ss'}$, which are constant in time. Given a current state $s_{\text{start}}$, the Gillespie algorithm works by identifying both the time $\tau$ and the final state $s_{\text{end}}$ of the following jump.

As a first step, the total rate $k_{\text{out}} = \sum_s k_{s s_{\text{start}}}$ of leaving state $s_{\text{start}}$ is computed. The time $\tau$ until the following jump is then drawn from an exponential distribution with mean $1/k_{\text{out}}$. The landing state is selected with probability $p(s) = k_{s s_{\text{start}}} / k_{\text{out}}$. The trajectory is thus constructed concatenating jumps.

First, the initial state $s_0$ is chosen (in our case, at random) at time $t=0$. A first jump $\left(\tau_1, s_1\right)$ is selected starting from $s_0$, and then a second $\left(\tau_2, s_2\right)$ starting from $s_1$. The process is repeated until one of two criteria is met, either a total time or a maximum number of steps. Average occupations can be computed considering that the system occupies state $s_i$ exactly for a time $\tau_i$ between jumps $i$ and $i+1$.

In our system, $s$ is a vector of $2N$ individual spins $s_i$ taking values in $\left\{-1,+1\right\}$. We will restrict the jumps to single spin flips. Given a state $s$, an input $x$ (external field) and a interaction matrix $W$, the transition where the $i${th} spin flips has a rate $k^{\left(i\right)}_{s} \propto e^{-\beta s_i h_i}$, with 
$h_i=\left(W s + x\right)_i$. The actual proportionality term (identical for all spins), which determines the time scale of the jumps, is not relevant since we are only interested in steady state properties and average occupancy.

To measure the average magnetization $m_x$ for each input $x$, we first select a random state $s_0$ and proceed to construct a trajectory up to a final time $T_{max}=5000$ or, alternatively, a maximum number of jumps $N_{max} = 10000$. The average magnetization of individual spins $m_x$ for that input is calculated after an initial transient time of $T_{transient} = 200$ is removed. 

Since we only consider the steady-state, we can evaluate the entropy production rate by summing the quantity $\Delta\sigma_n\equiv \log\frac{k_{s_{n+1}}^{\left(i\right)}}{k_{s_{n}}^{\left(i\right)}}=-2\beta s_{n,i}h_{n,i}$ for each jump $s_n\to s_{n+1}$, consisting of a single spin flip, and dividing by the total time~\cite{Loos_potts}.

\subsection{Task performance and parameter optimization.}

Given an input-output pair $\left(x^\mu, y^\mu\right)$ from the  set $\mathcal{D}=\left(x, y\right)$, we measure task performance by first computing $m_{x^\mu}$ and then the error between the prediction $\hat{y}^\mu$ of the final readout and the target $y^\mu$. We use two different readouts, with their respective loss functions:

\begin{itemize}
    \item \emph{Cross Entropy loss}: the "logit" vector $h^\mu=W_{out} m_{x^\mu}$ is passed through a Softmax function, thus getting the normalized estimated output probabilities $p^\mu_k=\frac{e^{h^\mu_k}}{\sum_{l=1}^K e^{h^\mu_l}}$, with $K$ the number of output labels. The loss function then amounts at computing the cross-entropy with the targets $y^\mu$: $L=-\frac{1}{M}\sum_{\mu=1}^M \log p^\mu_{y^\mu}$;
    \item \emph{MSE loss}: we compute the loss as $L=\frac{1}{2M} \sum_{\mu=1}^M \left( y^\mu - W_{out} m_{x^\mu}\right)^2$.
\end{itemize}

The minimization of a loss $L$ with respect to $W_{out}$ was performed either via a linear solver (for MSE) or a multinomial classifier solver (for CE), using standard libraries in julia, which retrieve optimal $W_{out}^*$ at fixed $W$, for the full input set.
We used MSE loss for the binary classification in the Random Task, whereas we employed the CE loss for multi-label classification in the MNIST-1D task.

\subsubsection{Optimization of $W$: SPSA}
Due to the stochastic nature of the dynamics, the optimization of the interaction parameters $W$ cannot be performed with standard gradient-based methods. Additionally, typical gradient evaluation through finite difference quickly becomes prohibitive as the number of independent parameters in $W$ grows.
To overcome this issue, we employ Simultaneous Perturbation Stochastic Approximation (SPSA)~\cite{spsa_multivariate,spsa_implementation}, where the gradient is approximated via a single finite difference in a random direction of the parameter space.

To evaluate the gradient $\nabla \mathcal{L}|_W$, a random vector $\delta W$ is constructed at every update step. Two symmetrical parameters configurations are constructed: $W^\pm = W \pm \delta W$. Independent dynamics are simulated to produce the average spin magnetizations $m^\pm$ and measure entropy production rates $\Sigma^\pm$. The average magnetizations  $m^\pm$ are thus used to compute the performance losses $\mathcal{G}^\pm$. Finally the gradient approximation reads $\nabla \mathcal{L}|_W \approx \left[\Sigma^+ - \Sigma^- + \alpha (\mathcal{G}^- - \mathcal{G}^+) \right] \frac{\delta W}{2|\delta W|}$. To avoid being trapped into local minima, we performed several initializations for each value of $\alpha$.

\section{Details on 2-spin system}

\subsection{Steady state}
The stationary state can be computed by imposing the stationary condition in Eq.~\ref{eq_sm:ctmc} and the normalization of $p$, thus getting~\cite{ngampruetikorn_energy_2020}:
\begin{equation}
p\left(s|x\right)=e^{-\beta \left(F + \delta F\right)} / Z,
\end{equation}
where
\begin{equation}
    F=x_1 s_1 + x_2 s_2 + J_s s_1 s_2
    \label{eq_sm:spin_pss_F}
\end{equation}
and
\begin{equation}
\delta F=-\beta^{-1}\log\left[e^{\beta J_a s_{1}s_{2}}\frac{\cosh\left(\beta\left(x_{1}-2 J_a s_{2}\right)\right)}{\cosh\beta x_{1}+\cosh\beta x_{2}}+e^{-\beta J_a s_{1}s_{2}}\frac{\cosh\left(\beta\left(x_{2}- 2 J_a s_{1}\right)\right)}{\cosh\beta x_{1}+\cosh\beta x_{2}}\right].
    \label{eq_sm:spin_pss_dF}
\end{equation}

\section{Computation-dissipation bottleneck in linear systems}
Let us consider a system whose dynamics, in the presence of a constant input $x$, is described by a multi-dimensional Ornstein-Uhlenbeck process:
\begin{equation}
\dot{s}=W s + x + \sigma_{s}\xi   
\end{equation}
with $\left\langle \xi\xi^{T}\right\rangle =\delta\left(t-t'\right)\mathcal{I}$, where $\mathcal{I}$ is the identity matrix. The (generally non-equilibrium) steady
state distribution ${p\left(s|x\right)}$ is a Gaussian with mean $m_x=W^{-1} x$ and whose covariance $C$
solves the Lyapunov equation:
\begin{equation}
W C +C W^{T} + \sigma_s^{2}\mathcal{I} = 0.
\label{eq_sm:lyapunov}
\end{equation}

Let us consider a noisy linear function $y=w_{0}^T x+\xi_{y}$,
with $\left\langle \xi_y\right\rangle = 0$ and $\sigma^2_{y}=\left\langle \xi^2_{y}\right\rangle $. Assuming $x$ is a Gaussian with mean zero and covariance $C_{x}$, one has $\left\langle y^2\right\rangle = w^T_{0}C_{x}w_{0}+\sigma^2_{y}$ and $C_{sy}=\left\langle sy\right\rangle =-W^{-1}\left\langle xy\right\rangle$.

To compute the mutual information, we use
\begin{equation}
I\left(s,y\right)=H\left(y\right)-H\left(y|s\right)
\end{equation}
and the relation for the entropy of a zero-mean, $d$ dimensional Gaussian variable $z$ with covariance $C_z$, $H\left(z\right)=\frac{1}{2}\log\left(\left(2\pi e\right)^{d}\det C_z\right)$, to get:
\begin{equation}
    I\left(s,y\right)=\frac{1}{2}\log\det\left( W^{-1}C_{x}W^{-T}+C\right)-\frac{1}{2}\log\det \left(W^{-1}C_{x|y}W^{-T}+C\right)
\end{equation}
where we used  that the covariance matrix $C_{s}=\left\langle ss^{T}\right\rangle$, averaged over the entire input distribution, equals $C_{s}=W^{-1}C_{x}W^{-T}+C$ and that $C_{s|y}=C_{s}-C_{sy}C_{y}^{-1}C_{ys}$, with $C_{s|y}$ the conditional covariance matrix of $s$ given $y$.

As shown in~\cite{seara_irreversibility_2021}, the entropy production in the presence of a given input $x$ can be computed in terms of an integral
\begin{equation}
\sigma=\int_{-\infty}^{+\infty}\frac{d\omega}{2\pi}\mathcal{E}\left(\omega\right)
\label{eq_sm:ep_density}
\end{equation}
where the density $\mathcal{E}\left(\omega\right)$ is given by:
\begin{equation}
\mathcal{E}\left(\omega\right)=\frac{1}{2} \Tr \left[C\left(\omega\right)\left(C^{-1}\left(-\omega\right)-C^{-1}\left(\omega\right)\right)\right],
\label{eq_sm:ep_total}
\end{equation}
with $C\left(\omega\right)$ the Fourier Transform of the steady state auto-correlation $C\left(t-t'\right)=\left<s\left(t\right) s^T\left(t'\right)\right>$.

The expressions derived thus far can be used to obtain the computation-dissipation bottleneck over any parametrization of the coupling matrix $W$ by numerical optimization, for different values of the tradeoff parameter $\alpha$. To exemplify the approach, the next section treats a 2-dimensional case where simple analytical expressions can be derived and a full enumeration of the parameter space is viable.

\subsection{\label{sec:cp2d}An example of a computation-dissipation bottleneck in a 2-dimensional case}
Let us then consider the case of a 2-particle system with an interaction matrix of the form:
\begin{equation}
W=\left(\begin{array}{cc}
-1 & J_s+J_a\\
J_s-J_a & -1
\end{array}\right). 
\end{equation}
Stability is guaranteed for $\Delta=1+J_a^{2}-J_s^{2}>0$.
The solution of the Lyapunov Eq.~\ref{eq_sm:lyapunov} for an input noise with covariance $\sigma_s^{2}\mathcal{I}$ is:
\begin{equation}
C=\frac{\sigma_s^{2}}{2\Delta}\left(\begin{array}{cc}
1+J_s J_a+J_a^{2} & J_s\\
J_s & 1-J_s J_a+J_a^{2}
\end{array}\right).
\end{equation}

The entropy production can be evaluated using Eq.~\ref{eq_sm:ep_density} and the Fourier transform of the system's Green function:
\begin{equation}
G\left(\omega\right)=\left(i\omega-W\right)^{-1}=\frac{1}{\Delta-\omega^{2}+2i\omega}\left(\begin{array}{cc}
1+i\omega & J_s+J_a\\
J_s-J_a & 1+i\omega
\end{array}\right).  
\end{equation}

From the Fourier Transform of the steady-state auto-correlation
$
C\left(\omega\right)=G\left(\omega\right)G^{\dagger}\left(\omega\right)
$
we get for the entropy production density:
\begin{equation}
\mathcal{E}\left(\omega\right)=\frac{8\omega^{2}J_a^{2}}{\left|\left(1+i\omega\right)^{2}+J_a^{2}-J_s^{2}\right|^{2}}.
\end{equation}

After integration in Eq.~\ref{eq_sm:ep_total}, and noting that $C$ doesn't depend on $x$, we get for a stable system:
\begin{equation}
\Sigma=2J_a^{2}.
\end{equation}

\begin{figure}
    \centering
    \includegraphics[width=0.85\linewidth]{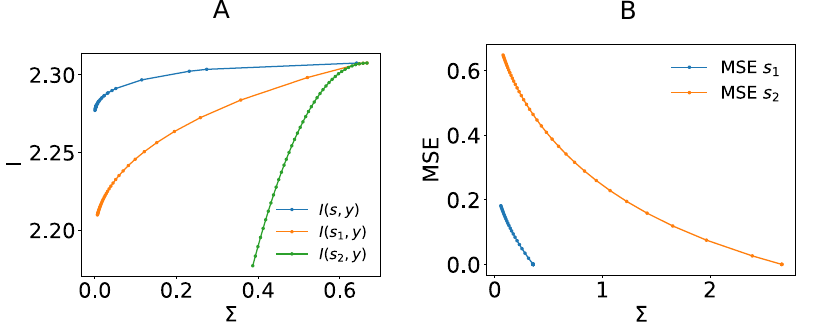}
    \caption{
    \textbf{A}: Optimal mutual information $I\left(s,y\right)$ between the system state $s$ and the output $y$ (blue curve) as a function of the entropy production rate at steady state $\Sigma$ in a 2-particle linear system described in the text. Parameters are specified in section~\ref{sec:cp2d}. We also show the single-particle mutual information $I\left(s_1,y\right)$ and $I\left(s_2,y\right)$ in the orange and green curve, respectively.
    \textbf{B}: Same as in \textbf{A} for the optimal squared deviation MSE when the output $y$ is linearly read out at each unit $s_i$, namely $\left<\left(y-s_i\right)^2\right>$.}
    \label{Fig:reservoir}
\end{figure}

We show in Fig.~\ref{Fig:reservoir} the results for a system with $\sigma_s=0.1$
tasked to compute a linear function $y=w_{0}^T x+\xi_{y}$ with $w_{0}=\left[\cos\left(\frac{\pi}{6}\right),\sin\left(\frac{\pi}{6}\right)\right]$, and $\xi_y$ a zero-mean Gaussian variable with standard deviation $\sigma_y=0.1$.

The trade-off between entropy production and output information is controlled by the degree of asymmetry in the entries of the vector $w_0$. In a similar vein, each particle $s_i$ can be used as a direct readout for the output $y$. In such a case, the average squared deviation $MSE_i = \left<\left(y-s_i\right)^2\right>$ at steady state again shows a characteristic front with respect to entropy production.

\bibliography{references}

\begin{thebibliography}{86}%
\makeatletter
\providecommand \@ifxundefined [1]{%
 \@ifx{#1\undefined}
}%
\providecommand \@ifnum [1]{%
 \ifnum #1\expandafter \@firstoftwo
 \else \expandafter \@secondoftwo
 \fi
}%
\providecommand \@ifx [1]{%
 \ifx #1\expandafter \@firstoftwo
 \else \expandafter \@secondoftwo
 \fi
}%
\providecommand \natexlab [1]{#1}%
\providecommand \enquote  [1]{``#1''}%
\providecommand \bibnamefont  [1]{#1}%
\providecommand \bibfnamefont [1]{#1}%
\providecommand \citenamefont [1]{#1}%
\providecommand \href@noop [0]{\@secondoftwo}%
\providecommand \href [0]{\begingroup \@sanitize@url \@href}%
\providecommand \@href[1]{\@@startlink{#1}\@@href}%
\providecommand \@@href[1]{\endgroup#1\@@endlink}%
\providecommand \@sanitize@url [0]{\catcode `\\12\catcode `\$12\catcode
  `\&12\catcode `\#12\catcode `\^12\catcode `\_12\catcode `\%12\relax}%
\providecommand \@@startlink[1]{}%
\providecommand \@@endlink[0]{}%
\providecommand \url  [0]{\begingroup\@sanitize@url \@url }%
\providecommand \@url [1]{\endgroup\@href {#1}{\urlprefix }}%
\providecommand \urlprefix  [0]{URL }%
\providecommand \Eprint [0]{\href }%
\providecommand \doibase [0]{http://dx.doi.org/}%
\providecommand \selectlanguage [0]{\@gobble}%
\providecommand \bibinfo  [0]{\@secondoftwo}%
\providecommand \bibfield  [0]{\@secondoftwo}%
\providecommand \translation [1]{[#1]}%
\providecommand \BibitemOpen [0]{}%
\providecommand \bibitemStop [0]{}%
\providecommand \bibitemNoStop [0]{.\EOS\space}%
\providecommand \EOS [0]{\spacefactor3000\relax}%
\providecommand \BibitemShut  [1]{\csname bibitem#1\endcsname}%
\let\auto@bib@innerbib\@empty
\bibitem [{\citenamefont {Seifert}(2005)}]{seifert_entropy_2005}%
  \BibitemOpen
  \bibfield  {author} {\bibinfo {author} {\bibfnamefont {U.}~\bibnamefont
  {Seifert}},\ }\href {\doibase 10.1103/PhysRevLett.95.040602} {\bibfield
  {journal} {\bibinfo  {journal} {Physical Review Letters}\ }\textbf {\bibinfo
  {volume} {95}},\ \bibinfo {pages} {040602} (\bibinfo {year} {2005})},\
  \bibinfo {note} {arXiv:cond-mat/0503686}\BibitemShut {NoStop}%
\bibitem [{\citenamefont {Seifert}(2012)}]{seifert_stochastic_2012}%
  \BibitemOpen
  \bibfield  {author} {\bibinfo {author} {\bibfnamefont {U.}~\bibnamefont
  {Seifert}},\ }\href {\doibase 10.1088/0034-4885/75/12/126001} {\bibfield
  {journal} {\bibinfo  {journal} {Reports on Progress in Physics}\ }\textbf
  {\bibinfo {volume} {75}},\ \bibinfo {pages} {126001} (\bibinfo {year}
  {2012})}\BibitemShut {NoStop}%
\bibitem [{\citenamefont {Van~den Broeck}\ and\ \citenamefont
  {Esposito}(2015)}]{van_den_broeck_ensemble_2015}%
  \BibitemOpen
  \bibfield  {author} {\bibinfo {author} {\bibfnamefont {C.}~\bibnamefont
  {Van~den Broeck}}\ and\ \bibinfo {author} {\bibfnamefont {M.}~\bibnamefont
  {Esposito}},\ }\href {\doibase 10.1016/j.physa.2014.04.035} {\bibfield
  {journal} {\bibinfo  {journal} {Physica A: Statistical Mechanics and its
  Applications}\ }\textbf {\bibinfo {volume} {418}},\ \bibinfo {pages} {6}
  (\bibinfo {year} {2015})}\BibitemShut {NoStop}%
\bibitem [{\citenamefont {Peliti}\ and\ \citenamefont
  {Pigolotti}(2021)}]{peliti_stochastic_2021}%
  \BibitemOpen
  \bibfield  {author} {\bibinfo {author} {\bibfnamefont {L.}~\bibnamefont
  {Peliti}}\ and\ \bibinfo {author} {\bibfnamefont {S.}~\bibnamefont
  {Pigolotti}},\ }\href@noop {} {\emph {\bibinfo {title} {Stochastic
  {Thermodynamics}: {An} {Introduction}}}}\ (\bibinfo  {publisher} {Princeton
  University Press},\ \bibinfo {year} {2021})\BibitemShut {NoStop}%
\bibitem [{\citenamefont {Andrieux}\ \emph {et~al.}(2007)\citenamefont
  {Andrieux}, \citenamefont {Gaspard}, \citenamefont {Ciliberto}, \citenamefont
  {Garnier}, \citenamefont {Joubaud},\ and\ \citenamefont
  {Petrosyan}}]{andrieux2007entropy}%
  \BibitemOpen
  \bibfield  {author} {\bibinfo {author} {\bibfnamefont {D.}~\bibnamefont
  {Andrieux}}, \bibinfo {author} {\bibfnamefont {P.}~\bibnamefont {Gaspard}},
  \bibinfo {author} {\bibfnamefont {S.}~\bibnamefont {Ciliberto}}, \bibinfo
  {author} {\bibfnamefont {N.}~\bibnamefont {Garnier}}, \bibinfo {author}
  {\bibfnamefont {S.}~\bibnamefont {Joubaud}}, \ and\ \bibinfo {author}
  {\bibfnamefont {A.}~\bibnamefont {Petrosyan}},\ }\href@noop {} {\bibfield
  {journal} {\bibinfo  {journal} {Physical review letters}\ }\textbf {\bibinfo
  {volume} {98}},\ \bibinfo {pages} {150601} (\bibinfo {year}
  {2007})}\BibitemShut {NoStop}%
\bibitem [{\citenamefont {Parrondo}\ \emph {et~al.}(2009)\citenamefont
  {Parrondo}, \citenamefont {Van~den Broeck},\ and\ \citenamefont
  {Kawai}}]{parrondo2009entropy}%
  \BibitemOpen
  \bibfield  {author} {\bibinfo {author} {\bibfnamefont {J.~M.}\ \bibnamefont
  {Parrondo}}, \bibinfo {author} {\bibfnamefont {C.}~\bibnamefont {Van~den
  Broeck}}, \ and\ \bibinfo {author} {\bibfnamefont {R.}~\bibnamefont
  {Kawai}},\ }\href@noop {} {\bibfield  {journal} {\bibinfo  {journal} {New
  Journal of Physics}\ }\textbf {\bibinfo {volume} {11}},\ \bibinfo {pages}
  {073008} (\bibinfo {year} {2009})}\BibitemShut {NoStop}%
\bibitem [{\citenamefont {Landauer}(1961)}]{Landauer_irreversibility}%
  \BibitemOpen
  \bibfield  {author} {\bibinfo {author} {\bibfnamefont {R.}~\bibnamefont
  {Landauer}},\ }\href {\doibase 10.1147/rd.53.0183} {\bibfield  {journal}
  {\bibinfo  {journal} {IBM Journal of Research and Development}\ }\textbf
  {\bibinfo {volume} {5}},\ \bibinfo {pages} {183} (\bibinfo {year}
  {1961})}\BibitemShut {NoStop}%
\bibitem [{\citenamefont {Bennett}(2003)}]{bennet_notes}%
  \BibitemOpen
  \bibfield  {author} {\bibinfo {author} {\bibfnamefont {C.~H.}\ \bibnamefont
  {Bennett}},\ }\href {\doibase https://doi.org/10.1016/S1355-2198(03)00039-X}
  {\bibfield  {journal} {\bibinfo  {journal} {Studies in History and Philosophy
  of Science Part B: Studies in History and Philosophy of Modern Physics}\
  }\textbf {\bibinfo {volume} {34}},\ \bibinfo {pages} {501} (\bibinfo {year}
  {2003})},\ \bibinfo {note} {quantum Information and Computation}\BibitemShut
  {NoStop}%
\bibitem [{\citenamefont {Esposito}\ and\ \citenamefont {Van~den
  Broeck}(2011)}]{esposito_second_2011}%
  \BibitemOpen
  \bibfield  {author} {\bibinfo {author} {\bibfnamefont {M.}~\bibnamefont
  {Esposito}}\ and\ \bibinfo {author} {\bibfnamefont {C.}~\bibnamefont {Van~den
  Broeck}},\ }\href {\doibase 10.1209/0295-5075/95/40004} {\bibfield  {journal}
  {\bibinfo  {journal} {EPL (Europhysics Letters)}\ }\textbf {\bibinfo {volume}
  {95}},\ \bibinfo {pages} {40004} (\bibinfo {year} {2011})}\BibitemShut
  {NoStop}%
\bibitem [{\citenamefont {Bérut}\ \emph {et~al.}(2012)\citenamefont {Bérut},
  \citenamefont {Arakelyan}, \citenamefont {Petrosyan}, \citenamefont
  {Ciliberto}, \citenamefont {Dillenschneider},\ and\ \citenamefont
  {Lutz}}]{berut_experimental_2012}%
  \BibitemOpen
  \bibfield  {author} {\bibinfo {author} {\bibfnamefont {A.}~\bibnamefont
  {Bérut}}, \bibinfo {author} {\bibfnamefont {A.}~\bibnamefont {Arakelyan}},
  \bibinfo {author} {\bibfnamefont {A.}~\bibnamefont {Petrosyan}}, \bibinfo
  {author} {\bibfnamefont {S.}~\bibnamefont {Ciliberto}}, \bibinfo {author}
  {\bibfnamefont {R.}~\bibnamefont {Dillenschneider}}, \ and\ \bibinfo {author}
  {\bibfnamefont {E.}~\bibnamefont {Lutz}},\ }\href {\doibase
  10.1038/nature10872} {\bibfield  {journal} {\bibinfo  {journal} {Nature}\
  }\textbf {\bibinfo {volume} {483}},\ \bibinfo {pages} {187} (\bibinfo {year}
  {2012})}\BibitemShut {NoStop}%
\bibitem [{\citenamefont {Parrondo}\ \emph {et~al.}(2015)\citenamefont
  {Parrondo}, \citenamefont {Horowitz},\ and\ \citenamefont
  {Sagawa}}]{parrondo_thermodynamics_2015}%
  \BibitemOpen
  \bibfield  {author} {\bibinfo {author} {\bibfnamefont {J.~M.~R.}\
  \bibnamefont {Parrondo}}, \bibinfo {author} {\bibfnamefont {J.~M.}\
  \bibnamefont {Horowitz}}, \ and\ \bibinfo {author} {\bibfnamefont
  {T.}~\bibnamefont {Sagawa}},\ }\href {\doibase 10.1038/nphys3230} {\bibfield
  {journal} {\bibinfo  {journal} {Nature Physics}\ }\textbf {\bibinfo {volume}
  {11}},\ \bibinfo {pages} {131} (\bibinfo {year} {2015})}\BibitemShut
  {NoStop}%
\bibitem [{\citenamefont {Barato}\ and\ \citenamefont
  {Seifert}(2015)}]{barato_thermodynamic_2015}%
  \BibitemOpen
  \bibfield  {author} {\bibinfo {author} {\bibfnamefont {A.~C.}\ \bibnamefont
  {Barato}}\ and\ \bibinfo {author} {\bibfnamefont {U.}~\bibnamefont
  {Seifert}},\ }\href {\doibase 10.1103/PhysRevLett.114.158101} {\bibfield
  {journal} {\bibinfo  {journal} {Physical Review Letters}\ }\textbf {\bibinfo
  {volume} {114}},\ \bibinfo {pages} {158101} (\bibinfo {year} {2015})},\
  \bibinfo {note} {arXiv:1502.05944 [cond-mat, physics:physics]}\BibitemShut
  {NoStop}%
\bibitem [{\citenamefont {Seifert}(2018)}]{seifert_stochastic_2018}%
  \BibitemOpen
  \bibfield  {author} {\bibinfo {author} {\bibfnamefont {U.}~\bibnamefont
  {Seifert}},\ }\href {\doibase 10.1016/j.physa.2017.10.024} {\bibfield
  {journal} {\bibinfo  {journal} {Physica A: Statistical Mechanics and its
  Applications}\ }\textbf {\bibinfo {volume} {504}},\ \bibinfo {pages} {176}
  (\bibinfo {year} {2018})}\BibitemShut {NoStop}%
\bibitem [{\citenamefont {Koyuk}\ and\ \citenamefont
  {Seifert}(2020)}]{koyuk_thermodynamic_2020}%
  \BibitemOpen
  \bibfield  {author} {\bibinfo {author} {\bibfnamefont {T.}~\bibnamefont
  {Koyuk}}\ and\ \bibinfo {author} {\bibfnamefont {U.}~\bibnamefont
  {Seifert}},\ }\href {\doibase 10.1103/PhysRevLett.125.260604} {\bibfield
  {journal} {\bibinfo  {journal} {Physical Review Letters}\ }\textbf {\bibinfo
  {volume} {125}},\ \bibinfo {pages} {260604} (\bibinfo {year}
  {2020})}\BibitemShut {NoStop}%
\bibitem [{\citenamefont {Horowitz}\ and\ \citenamefont
  {Gingrich}(2020)}]{horowitz2020thermodynamic}%
  \BibitemOpen
  \bibfield  {author} {\bibinfo {author} {\bibfnamefont {J.~M.}\ \bibnamefont
  {Horowitz}}\ and\ \bibinfo {author} {\bibfnamefont {T.~R.}\ \bibnamefont
  {Gingrich}},\ }\href@noop {} {\bibfield  {journal} {\bibinfo  {journal}
  {Nature Physics}\ }\textbf {\bibinfo {volume} {16}},\ \bibinfo {pages} {15}
  (\bibinfo {year} {2020})}\BibitemShut {NoStop}%
\bibitem [{\citenamefont {Lan}\ \emph {et~al.}(2012)\citenamefont {Lan},
  \citenamefont {Sartori}, \citenamefont {Neumann}, \citenamefont {Sourjik},\
  and\ \citenamefont {Tu}}]{lan2012energy}%
  \BibitemOpen
  \bibfield  {author} {\bibinfo {author} {\bibfnamefont {G.}~\bibnamefont
  {Lan}}, \bibinfo {author} {\bibfnamefont {P.}~\bibnamefont {Sartori}},
  \bibinfo {author} {\bibfnamefont {S.}~\bibnamefont {Neumann}}, \bibinfo
  {author} {\bibfnamefont {V.}~\bibnamefont {Sourjik}}, \ and\ \bibinfo
  {author} {\bibfnamefont {Y.}~\bibnamefont {Tu}},\ }\href@noop {} {\bibfield
  {journal} {\bibinfo  {journal} {Nature physics}\ }\textbf {\bibinfo {volume}
  {8}},\ \bibinfo {pages} {422} (\bibinfo {year} {2012})}\BibitemShut {NoStop}%
\bibitem [{\citenamefont {Sartori}\ \emph {et~al.}(2014)\citenamefont
  {Sartori}, \citenamefont {Granger}, \citenamefont {Lee},\ and\ \citenamefont
  {Horowitz}}]{sartori_thermodynamic_2014}%
  \BibitemOpen
  \bibfield  {author} {\bibinfo {author} {\bibfnamefont {P.}~\bibnamefont
  {Sartori}}, \bibinfo {author} {\bibfnamefont {L.}~\bibnamefont {Granger}},
  \bibinfo {author} {\bibfnamefont {C.~F.}\ \bibnamefont {Lee}}, \ and\
  \bibinfo {author} {\bibfnamefont {J.~M.}\ \bibnamefont {Horowitz}},\ }\href
  {\doibase 10.1371/journal.pcbi.1003974} {\bibfield  {journal} {\bibinfo
  {journal} {PLoS Computational Biology}\ }\textbf {\bibinfo {volume} {10}},\
  \bibinfo {pages} {e1003974} (\bibinfo {year} {2014})}\BibitemShut {NoStop}%
\bibitem [{\citenamefont {Ngampruetikorn}\ \emph {et~al.}(2020)\citenamefont
  {Ngampruetikorn}, \citenamefont {Schwab},\ and\ \citenamefont
  {Stephens}}]{ngampruetikorn_energy_2020}%
  \BibitemOpen
  \bibfield  {author} {\bibinfo {author} {\bibfnamefont {V.}~\bibnamefont
  {Ngampruetikorn}}, \bibinfo {author} {\bibfnamefont {D.~J.}\ \bibnamefont
  {Schwab}}, \ and\ \bibinfo {author} {\bibfnamefont {G.~J.}\ \bibnamefont
  {Stephens}},\ }\href {\doibase 10.1038/s41467-020-14806-y} {\bibfield
  {journal} {\bibinfo  {journal} {Nature Communications}\ }\textbf {\bibinfo
  {volume} {11}},\ \bibinfo {pages} {975} (\bibinfo {year} {2020})}\BibitemShut
  {NoStop}%
\bibitem [{\citenamefont {Sartori}\ and\ \citenamefont
  {Pigolotti}(2015)}]{sartori_thermodynamics_2015}%
  \BibitemOpen
  \bibfield  {author} {\bibinfo {author} {\bibfnamefont {P.}~\bibnamefont
  {Sartori}}\ and\ \bibinfo {author} {\bibfnamefont {S.}~\bibnamefont
  {Pigolotti}},\ }\href {\doibase 10.1103/PhysRevX.5.041039} {\bibfield
  {journal} {\bibinfo  {journal} {Physical Review X}\ }\textbf {\bibinfo
  {volume} {5}},\ \bibinfo {pages} {041039} (\bibinfo {year}
  {2015})}\BibitemShut {NoStop}%
\bibitem [{\citenamefont {Rao}\ and\ \citenamefont
  {Esposito}(2016)}]{espositoReaction2016}%
  \BibitemOpen
  \bibfield  {author} {\bibinfo {author} {\bibfnamefont {R.}~\bibnamefont
  {Rao}}\ and\ \bibinfo {author} {\bibfnamefont {M.}~\bibnamefont {Esposito}},\
  }\href {\doibase 10.1103/PhysRevX.6.041064} {\bibfield  {journal} {\bibinfo
  {journal} {Phys. Rev. X}\ }\textbf {\bibinfo {volume} {6}},\ \bibinfo {pages}
  {041064} (\bibinfo {year} {2016})}\BibitemShut {NoStop}%
\bibitem [{\citenamefont {Fritz}\ \emph {et~al.}(2020)\citenamefont {Fritz},
  \citenamefont {Nguyen},\ and\ \citenamefont
  {Seifert}}]{fritz_stochastic_2020}%
  \BibitemOpen
  \bibfield  {author} {\bibinfo {author} {\bibfnamefont {J.~H.}\ \bibnamefont
  {Fritz}}, \bibinfo {author} {\bibfnamefont {B.}~\bibnamefont {Nguyen}}, \
  and\ \bibinfo {author} {\bibfnamefont {U.}~\bibnamefont {Seifert}},\ }\href
  {\doibase 10.1063/5.0006115} {\bibfield  {journal} {\bibinfo  {journal} {The
  Journal of Chemical Physics}\ }\textbf {\bibinfo {volume} {152}},\ \bibinfo
  {pages} {235101} (\bibinfo {year} {2020})}\BibitemShut {NoStop}%
\bibitem [{\citenamefont {Karbowski}(2019)}]{karbowski_metabolic_2019}%
  \BibitemOpen
  \bibfield  {author} {\bibinfo {author} {\bibfnamefont {J.}~\bibnamefont
  {Karbowski}},\ }\href {\doibase 10.1152/jn.00092.2019} {\bibfield  {journal}
  {\bibinfo  {journal} {Journal of Neurophysiology}\ }\textbf {\bibinfo
  {volume} {122}},\ \bibinfo {pages} {1473} (\bibinfo {year}
  {2019})}\BibitemShut {NoStop}%
\bibitem [{\citenamefont {Barato}\ \emph {et~al.}(2013)\citenamefont {Barato},
  \citenamefont {Hartich},\ and\ \citenamefont
  {Seifert}}]{barato2013information}%
  \BibitemOpen
  \bibfield  {author} {\bibinfo {author} {\bibfnamefont {A.}~\bibnamefont
  {Barato}}, \bibinfo {author} {\bibfnamefont {D.}~\bibnamefont {Hartich}}, \
  and\ \bibinfo {author} {\bibfnamefont {U.}~\bibnamefont {Seifert}},\
  }\href@noop {} {\bibfield  {journal} {\bibinfo  {journal} {Physical Review
  E}\ }\textbf {\bibinfo {volume} {87}},\ \bibinfo {pages} {042104} (\bibinfo
  {year} {2013})}\BibitemShut {NoStop}%
\bibitem [{\citenamefont {Barato}\ \emph {et~al.}(2014)\citenamefont {Barato},
  \citenamefont {Hartich},\ and\ \citenamefont
  {Seifert}}]{barato2014efficiency}%
  \BibitemOpen
  \bibfield  {author} {\bibinfo {author} {\bibfnamefont {A.~C.}\ \bibnamefont
  {Barato}}, \bibinfo {author} {\bibfnamefont {D.}~\bibnamefont {Hartich}}, \
  and\ \bibinfo {author} {\bibfnamefont {U.}~\bibnamefont {Seifert}},\
  }\href@noop {} {\bibfield  {journal} {\bibinfo  {journal} {New Journal of
  Physics}\ }\textbf {\bibinfo {volume} {16}},\ \bibinfo {pages} {103024}
  (\bibinfo {year} {2014})}\BibitemShut {NoStop}%
\bibitem [{\citenamefont {Herpich}\ \emph {et~al.}(2018)\citenamefont
  {Herpich}, \citenamefont {Thingna},\ and\ \citenamefont
  {Esposito}}]{herpich_collective_2018}%
  \BibitemOpen
  \bibfield  {author} {\bibinfo {author} {\bibfnamefont {T.}~\bibnamefont
  {Herpich}}, \bibinfo {author} {\bibfnamefont {J.}~\bibnamefont {Thingna}}, \
  and\ \bibinfo {author} {\bibfnamefont {M.}~\bibnamefont {Esposito}},\ }\href
  {\doibase 10.1103/PhysRevX.8.031056} {\bibfield  {journal} {\bibinfo
  {journal} {Physical Review X}\ }\textbf {\bibinfo {volume} {8}},\ \bibinfo
  {pages} {031056} (\bibinfo {year} {2018})}\BibitemShut {NoStop}%
\bibitem [{\citenamefont {Suñé}\ and\ \citenamefont
  {Imparato}(2019)}]{sune_out--equilibrium_2019}%
  \BibitemOpen
  \bibfield  {author} {\bibinfo {author} {\bibfnamefont {M.}~\bibnamefont
  {Suñé}}\ and\ \bibinfo {author} {\bibfnamefont {A.}~\bibnamefont
  {Imparato}},\ }\href {\doibase 10.1103/PhysRevLett.123.070601} {\bibfield
  {journal} {\bibinfo  {journal} {Physical Review Letters}\ }\textbf {\bibinfo
  {volume} {123}},\ \bibinfo {pages} {070601} (\bibinfo {year}
  {2019})}\BibitemShut {NoStop}%
\bibitem [{\citenamefont {Herpich}\ \emph {et~al.}(2020)\citenamefont
  {Herpich}, \citenamefont {Cossetto}, \citenamefont {Falasco},\ and\
  \citenamefont {Esposito}}]{herpich_stochastic_2020-1}%
  \BibitemOpen
  \bibfield  {author} {\bibinfo {author} {\bibfnamefont {T.}~\bibnamefont
  {Herpich}}, \bibinfo {author} {\bibfnamefont {T.}~\bibnamefont {Cossetto}},
  \bibinfo {author} {\bibfnamefont {G.}~\bibnamefont {Falasco}}, \ and\
  \bibinfo {author} {\bibfnamefont {M.}~\bibnamefont {Esposito}},\ }\href
  {\doibase 10.1088/1367-2630/ab882f} {\bibfield  {journal} {\bibinfo
  {journal} {New Journal of Physics}\ }\textbf {\bibinfo {volume} {22}},\
  \bibinfo {pages} {063005} (\bibinfo {year} {2020})}\BibitemShut {NoStop}%
\bibitem [{\citenamefont {Cofré}\ and\ \citenamefont
  {Maldonado}(2018)}]{cofre_information_2018}%
  \BibitemOpen
  \bibfield  {author} {\bibinfo {author} {\bibfnamefont {R.}~\bibnamefont
  {Cofré}}\ and\ \bibinfo {author} {\bibfnamefont {C.}~\bibnamefont
  {Maldonado}},\ }\href {\doibase 10.3390/e20010034} {\bibfield  {journal}
  {\bibinfo  {journal} {Entropy}\ }\textbf {\bibinfo {volume} {20}},\ \bibinfo
  {pages} {34} (\bibinfo {year} {2018})}\BibitemShut {NoStop}%
\bibitem [{\citenamefont {Cofré}\ \emph {et~al.}(2019)\citenamefont {Cofré},
  \citenamefont {Videla},\ and\ \citenamefont
  {Rosas}}]{cofre_introduction_2019}%
  \BibitemOpen
  \bibfield  {author} {\bibinfo {author} {\bibfnamefont {R.}~\bibnamefont
  {Cofré}}, \bibinfo {author} {\bibfnamefont {L.}~\bibnamefont {Videla}}, \
  and\ \bibinfo {author} {\bibfnamefont {F.}~\bibnamefont {Rosas}},\ }\href
  {\doibase 10.3390/e21090884} {\bibfield  {journal} {\bibinfo  {journal}
  {Entropy}\ }\textbf {\bibinfo {volume} {21}},\ \bibinfo {pages} {884}
  (\bibinfo {year} {2019})}\BibitemShut {NoStop}%
\bibitem [{\citenamefont {Lynn}\ \emph
  {et~al.}(2022{\natexlab{a}})\citenamefont {Lynn}, \citenamefont {Holmes},
  \citenamefont {Bialek},\ and\ \citenamefont {Schwab}}]{lynn_emergence_2022}%
  \BibitemOpen
  \bibfield  {author} {\bibinfo {author} {\bibfnamefont {C.~W.}\ \bibnamefont
  {Lynn}}, \bibinfo {author} {\bibfnamefont {C.~M.}\ \bibnamefont {Holmes}},
  \bibinfo {author} {\bibfnamefont {W.}~\bibnamefont {Bialek}}, \ and\ \bibinfo
  {author} {\bibfnamefont {D.~J.}\ \bibnamefont {Schwab}},\ }\href {\doibase
  10.1103/PhysRevE.106.034102} {\bibfield  {journal} {\bibinfo  {journal}
  {Physical Review E}\ }\textbf {\bibinfo {volume} {106}},\ \bibinfo {pages}
  {034102} (\bibinfo {year} {2022}{\natexlab{a}})}\BibitemShut {NoStop}%
\bibitem [{\citenamefont {Lynn}\ \emph
  {et~al.}(2022{\natexlab{b}})\citenamefont {Lynn}, \citenamefont {Holmes},
  \citenamefont {Bialek},\ and\ \citenamefont
  {Schwab}}]{lynn_decomposing_2022}%
  \BibitemOpen
  \bibfield  {author} {\bibinfo {author} {\bibfnamefont {C.~W.}\ \bibnamefont
  {Lynn}}, \bibinfo {author} {\bibfnamefont {C.~M.}\ \bibnamefont {Holmes}},
  \bibinfo {author} {\bibfnamefont {W.}~\bibnamefont {Bialek}}, \ and\ \bibinfo
  {author} {\bibfnamefont {D.~J.}\ \bibnamefont {Schwab}},\ }\href {\doibase
  10.1103/PhysRevLett.129.118101} {\bibfield  {journal} {\bibinfo  {journal}
  {Physical Review Letters}\ }\textbf {\bibinfo {volume} {129}},\ \bibinfo
  {pages} {118101} (\bibinfo {year} {2022}{\natexlab{b}})}\BibitemShut
  {NoStop}%
\bibitem [{\citenamefont {Wolpert}\ \emph {et~al.}(2019)\citenamefont
  {Wolpert}, \citenamefont {Kolchinsky},\ and\ \citenamefont
  {Owen}}]{wolpert_spacetime_2019}%
  \BibitemOpen
  \bibfield  {author} {\bibinfo {author} {\bibfnamefont {D.~H.}\ \bibnamefont
  {Wolpert}}, \bibinfo {author} {\bibfnamefont {A.}~\bibnamefont {Kolchinsky}},
  \ and\ \bibinfo {author} {\bibfnamefont {J.~A.}\ \bibnamefont {Owen}},\
  }\href {\doibase 10.1038/s41467-019-09542-x} {\bibfield  {journal} {\bibinfo
  {journal} {Nature Communications}\ }\textbf {\bibinfo {volume} {10}},\
  \bibinfo {pages} {1727} (\bibinfo {year} {2019})}\BibitemShut {NoStop}%
\bibitem [{\citenamefont {Wolpert}(2019)}]{wolpert_stochastic_2019}%
  \BibitemOpen
  \bibfield  {author} {\bibinfo {author} {\bibfnamefont {D.~H.}\ \bibnamefont
  {Wolpert}},\ }\href {\doibase 10.1088/1751-8121/ab0850} {\bibfield  {journal}
  {\bibinfo  {journal} {Journal of Physics A: Mathematical and Theoretical}\
  }\textbf {\bibinfo {volume} {52}},\ \bibinfo {pages} {193001} (\bibinfo
  {year} {2019})}\BibitemShut {NoStop}%
\bibitem [{\citenamefont {Wolpert}\ and\ \citenamefont
  {Kolchinsky}(2020)}]{wolpert_thermodynamics_2020}%
  \BibitemOpen
  \bibfield  {author} {\bibinfo {author} {\bibfnamefont {D.~H.}\ \bibnamefont
  {Wolpert}}\ and\ \bibinfo {author} {\bibfnamefont {A.}~\bibnamefont
  {Kolchinsky}},\ }\href {\doibase 10.1088/1367-2630/ab82b8} {\bibfield
  {journal} {\bibinfo  {journal} {New Journal of Physics}\ }\textbf {\bibinfo
  {volume} {22}},\ \bibinfo {pages} {063047} (\bibinfo {year}
  {2020})}\BibitemShut {NoStop}%
\bibitem [{\citenamefont {Ivlev}\ \emph {et~al.}(2015)\citenamefont {Ivlev},
  \citenamefont {Bartnick}, \citenamefont {Heinen}, \citenamefont {Du},
  \citenamefont {Nosenko},\ and\ \citenamefont
  {L{\"o}wen}}]{ivlev2015statistical}%
  \BibitemOpen
  \bibfield  {author} {\bibinfo {author} {\bibfnamefont {A.~V.}\ \bibnamefont
  {Ivlev}}, \bibinfo {author} {\bibfnamefont {J.}~\bibnamefont {Bartnick}},
  \bibinfo {author} {\bibfnamefont {M.}~\bibnamefont {Heinen}}, \bibinfo
  {author} {\bibfnamefont {C.-R.}\ \bibnamefont {Du}}, \bibinfo {author}
  {\bibfnamefont {V.}~\bibnamefont {Nosenko}}, \ and\ \bibinfo {author}
  {\bibfnamefont {H.}~\bibnamefont {L{\"o}wen}},\ }\href@noop {} {\bibfield
  {journal} {\bibinfo  {journal} {Physical Review X}\ }\textbf {\bibinfo
  {volume} {5}},\ \bibinfo {pages} {011035} (\bibinfo {year}
  {2015})}\BibitemShut {NoStop}%
\bibitem [{\citenamefont {Crisanti}\ and\ \citenamefont
  {Sompolinsky}(1987)}]{crisanti_dynamics_1987}%
  \BibitemOpen
  \bibfield  {author} {\bibinfo {author} {\bibfnamefont {A.}~\bibnamefont
  {Crisanti}}\ and\ \bibinfo {author} {\bibfnamefont {H.}~\bibnamefont
  {Sompolinsky}},\ }\href {\doibase 10.1103/PhysRevA.36.4922} {\bibfield
  {journal} {\bibinfo  {journal} {Physical Review A}\ }\textbf {\bibinfo
  {volume} {36}},\ \bibinfo {pages} {4922} (\bibinfo {year}
  {1987})}\BibitemShut {NoStop}%
\bibitem [{\citenamefont {Crisanti}\ and\ \citenamefont
  {Sompolinsky}(1988)}]{crisanti_dynamics_1988}%
  \BibitemOpen
  \bibfield  {author} {\bibinfo {author} {\bibfnamefont {A.}~\bibnamefont
  {Crisanti}}\ and\ \bibinfo {author} {\bibfnamefont {H.}~\bibnamefont
  {Sompolinsky}},\ }\href {\doibase 10.1103/PhysRevA.37.4865} {\bibfield
  {journal} {\bibinfo  {journal} {Physical Review A}\ }\textbf {\bibinfo
  {volume} {37}},\ \bibinfo {pages} {4865} (\bibinfo {year}
  {1988})}\BibitemShut {NoStop}%
\bibitem [{\citenamefont {Aguilera}\ \emph {et~al.}(2021)\citenamefont
  {Aguilera}, \citenamefont {Moosavi},\ and\ \citenamefont
  {Shimazaki}}]{aguilera_unifying_2021}%
  \BibitemOpen
  \bibfield  {author} {\bibinfo {author} {\bibfnamefont {M.}~\bibnamefont
  {Aguilera}}, \bibinfo {author} {\bibfnamefont {S.~A.}\ \bibnamefont
  {Moosavi}}, \ and\ \bibinfo {author} {\bibfnamefont {H.}~\bibnamefont
  {Shimazaki}},\ }\href {\doibase 10.1038/s41467-021-20890-5} {\bibfield
  {journal} {\bibinfo  {journal} {Nature Communications}\ }\textbf {\bibinfo
  {volume} {12}},\ \bibinfo {pages} {1197} (\bibinfo {year}
  {2021})}\BibitemShut {NoStop}%
\bibitem [{\citenamefont {Aguilera}\ \emph {et~al.}(2022)\citenamefont
  {Aguilera}, \citenamefont {Igarashi},\ and\ \citenamefont
  {Shimazaki}}]{aguilera_nonequilibrium_2022}%
  \BibitemOpen
  \bibfield  {author} {\bibinfo {author} {\bibfnamefont {M.}~\bibnamefont
  {Aguilera}}, \bibinfo {author} {\bibfnamefont {M.}~\bibnamefont {Igarashi}},
  \ and\ \bibinfo {author} {\bibfnamefont {H.}~\bibnamefont {Shimazaki}},\
  }\href {http://arxiv.org/abs/2205.09886} {\enquote {\bibinfo {title}
  {Nonequilibrium thermodynamics of the asymmetric {Sherrington}-{Kirkpatrick}
  model},}\ } (\bibinfo {year} {2022}),\ \bibinfo {note} {arXiv:2205.09886
  [cond-mat]}\BibitemShut {NoStop}%
\bibitem [{\citenamefont {Ginzburg}\ and\ \citenamefont
  {Sompolinsky}(1994)}]{ginzburg_theory_1994}%
  \BibitemOpen
  \bibfield  {author} {\bibinfo {author} {\bibfnamefont {I.}~\bibnamefont
  {Ginzburg}}\ and\ \bibinfo {author} {\bibfnamefont {H.}~\bibnamefont
  {Sompolinsky}},\ }\href {\doibase 10.1103/PhysRevE.50.3171} {\bibfield
  {journal} {\bibinfo  {journal} {Physical Review E}\ }\textbf {\bibinfo
  {volume} {50}},\ \bibinfo {pages} {3171} (\bibinfo {year}
  {1994})}\BibitemShut {NoStop}%
\bibitem [{\citenamefont {Renart}\ \emph {et~al.}(2010)\citenamefont {Renart},
  \citenamefont {de~la Rocha}, \citenamefont {Bartho}, \citenamefont
  {Hollender}, \citenamefont {Parga}, \citenamefont {Reyes},\ and\
  \citenamefont {Harris}}]{renart_asynchronous_2010}%
  \BibitemOpen
  \bibfield  {author} {\bibinfo {author} {\bibfnamefont {A.}~\bibnamefont
  {Renart}}, \bibinfo {author} {\bibfnamefont {J.}~\bibnamefont {de~la Rocha}},
  \bibinfo {author} {\bibfnamefont {P.}~\bibnamefont {Bartho}}, \bibinfo
  {author} {\bibfnamefont {L.}~\bibnamefont {Hollender}}, \bibinfo {author}
  {\bibfnamefont {N.}~\bibnamefont {Parga}}, \bibinfo {author} {\bibfnamefont
  {A.}~\bibnamefont {Reyes}}, \ and\ \bibinfo {author} {\bibfnamefont {K.~D.}\
  \bibnamefont {Harris}},\ }\href {\doibase 10.1126/science.1179850} {\bibfield
   {journal} {\bibinfo  {journal} {Science}\ }\textbf {\bibinfo {volume}
  {327}},\ \bibinfo {pages} {587} (\bibinfo {year} {2010})}\BibitemShut
  {NoStop}%
\bibitem [{\citenamefont {Roudi}\ \emph {et~al.}(2015)\citenamefont {Roudi},
  \citenamefont {Dunn},\ and\ \citenamefont
  {Hertz}}]{roudi_multi-neuronal_2015}%
  \BibitemOpen
  \bibfield  {author} {\bibinfo {author} {\bibfnamefont {Y.}~\bibnamefont
  {Roudi}}, \bibinfo {author} {\bibfnamefont {B.}~\bibnamefont {Dunn}}, \ and\
  \bibinfo {author} {\bibfnamefont {J.}~\bibnamefont {Hertz}},\ }\href
  {\doibase 10.1016/j.conb.2014.10.011} {\bibfield  {journal} {\bibinfo
  {journal} {Current Opinion in Neurobiology}\ }\textbf {\bibinfo {volume}
  {32}},\ \bibinfo {pages} {38} (\bibinfo {year} {2015})}\BibitemShut {NoStop}%
\bibitem [{\citenamefont {Dunn}\ \emph {et~al.}(2015)\citenamefont {Dunn},
  \citenamefont {Mørreaunet},\ and\ \citenamefont
  {Roudi}}]{dunn_correlations_2015}%
  \BibitemOpen
  \bibfield  {author} {\bibinfo {author} {\bibfnamefont {B.}~\bibnamefont
  {Dunn}}, \bibinfo {author} {\bibfnamefont {M.}~\bibnamefont {Mørreaunet}}, \
  and\ \bibinfo {author} {\bibfnamefont {Y.}~\bibnamefont {Roudi}},\ }\href
  {\doibase 10.1371/journal.pcbi.1004052} {\bibfield  {journal} {\bibinfo
  {journal} {PLOS Computational Biology}\ }\textbf {\bibinfo {volume} {11}},\
  \bibinfo {pages} {e1004052} (\bibinfo {year} {2015})}\BibitemShut {NoStop}%
\bibitem [{\citenamefont {Shi}\ \emph {et~al.}(2023)\citenamefont {Shi},
  \citenamefont {Zeraati}, \citenamefont {Levina},\ and\ \citenamefont
  {Engel}}]{shi_spatial_2022}%
  \BibitemOpen
  \bibfield  {author} {\bibinfo {author} {\bibfnamefont {Y.-L.}\ \bibnamefont
  {Shi}}, \bibinfo {author} {\bibfnamefont {R.}~\bibnamefont {Zeraati}},
  \bibinfo {author} {\bibfnamefont {A.}~\bibnamefont {Levina}}, \ and\ \bibinfo
  {author} {\bibfnamefont {T.~A.}\ \bibnamefont {Engel}},\ }\href@noop {}
  {\bibfield  {journal} {\bibinfo  {journal} {Physical Review Research}\
  }\textbf {\bibinfo {volume} {5}},\ \bibinfo {pages} {013005} (\bibinfo {year}
  {2023})}\BibitemShut {NoStop}%
\bibitem [{\citenamefont {Loos}\ and\ \citenamefont {Klapp}(2020)}]{Loos_2020}%
  \BibitemOpen
  \bibfield  {author} {\bibinfo {author} {\bibfnamefont {S.~A.~M.}\
  \bibnamefont {Loos}}\ and\ \bibinfo {author} {\bibfnamefont {S.~H.~L.}\
  \bibnamefont {Klapp}},\ }\href {\doibase 10.1088/1367-2630/abcc1e} {\bibfield
   {journal} {\bibinfo  {journal} {New Journal of Physics}\ }\textbf {\bibinfo
  {volume} {22}},\ \bibinfo {pages} {123051} (\bibinfo {year}
  {2020})}\BibitemShut {NoStop}%
\bibitem [{\citenamefont {Loos}\ \emph {et~al.}(2023)\citenamefont {Loos},
  \citenamefont {Arabha}, \citenamefont {Rajabpour}, \citenamefont
  {Hassanali},\ and\ \citenamefont {Rold{\'a}n}}]{Loos2023}%
  \BibitemOpen
  \bibfield  {author} {\bibinfo {author} {\bibfnamefont {S.~A.~M.}\
  \bibnamefont {Loos}}, \bibinfo {author} {\bibfnamefont {S.}~\bibnamefont
  {Arabha}}, \bibinfo {author} {\bibfnamefont {A.}~\bibnamefont {Rajabpour}},
  \bibinfo {author} {\bibfnamefont {A.}~\bibnamefont {Hassanali}}, \ and\
  \bibinfo {author} {\bibfnamefont {{\'E}.}~\bibnamefont {Rold{\'a}n}},\ }\href
  {\doibase 10.1038/s41598-023-31583-y} {\bibfield  {journal} {\bibinfo
  {journal} {Scientific Reports}\ }\textbf {\bibinfo {volume} {13}},\ \bibinfo
  {pages} {4517} (\bibinfo {year} {2023})}\BibitemShut {NoStop}%
\bibitem [{\citenamefont {Schnakenberg}(1976)}]{Schnakenberg_formula}%
  \BibitemOpen
  \bibfield  {author} {\bibinfo {author} {\bibfnamefont {J.}~\bibnamefont
  {Schnakenberg}},\ }\href {\doibase 10.1103/RevModPhys.48.571} {\bibfield
  {journal} {\bibinfo  {journal} {Rev. Mod. Phys.}\ }\textbf {\bibinfo {volume}
  {48}},\ \bibinfo {pages} {571} (\bibinfo {year} {1976})}\BibitemShut
  {NoStop}%
\bibitem [{\citenamefont {Rold\'an}\ and\ \citenamefont
  {Parrondo}(2010)}]{RoldanEstimating2010}%
  \BibitemOpen
  \bibfield  {author} {\bibinfo {author} {\bibfnamefont {E.}~\bibnamefont
  {Rold\'an}}\ and\ \bibinfo {author} {\bibfnamefont {J.~M.~R.}\ \bibnamefont
  {Parrondo}},\ }\href {\doibase 10.1103/PhysRevLett.105.150607} {\bibfield
  {journal} {\bibinfo  {journal} {Phys. Rev. Lett.}\ }\textbf {\bibinfo
  {volume} {105}},\ \bibinfo {pages} {150607} (\bibinfo {year}
  {2010})}\BibitemShut {NoStop}%
\bibitem [{\citenamefont {Bai}\ \emph {et~al.}(2019)\citenamefont {Bai},
  \citenamefont {Kolter},\ and\ \citenamefont {Koltun}}]{Bai_deep}%
  \BibitemOpen
  \bibfield  {author} {\bibinfo {author} {\bibfnamefont {S.}~\bibnamefont
  {Bai}}, \bibinfo {author} {\bibfnamefont {J.~Z.}\ \bibnamefont {Kolter}}, \
  and\ \bibinfo {author} {\bibfnamefont {V.}~\bibnamefont {Koltun}},\ }in\
  \href
  {https://proceedings.neurips.cc/paper/2019/file/01386bd6d8e091c2ab4c7c7de644d37b-Paper.pdf}
  {\emph {\bibinfo {booktitle} {Advances in Neural Information Processing
  Systems}}},\ Vol.~\bibinfo {volume} {32},\ \bibinfo {editor} {edited by\
  \bibinfo {editor} {\bibfnamefont {H.}~\bibnamefont {Wallach}}, \bibinfo
  {editor} {\bibfnamefont {H.}~\bibnamefont {Larochelle}}, \bibinfo {editor}
  {\bibfnamefont {A.}~\bibnamefont {Beygelzimer}}, \bibinfo {editor}
  {\bibfnamefont {F.}~\bibnamefont {d\textquotesingle Alch\'{e}-Buc}}, \bibinfo
  {editor} {\bibfnamefont {E.}~\bibnamefont {Fox}}, \ and\ \bibinfo {editor}
  {\bibfnamefont {R.}~\bibnamefont {Garnett}}}\ (\bibinfo  {publisher} {Curran
  Associates, Inc.},\ \bibinfo {year} {2019})\BibitemShut {NoStop}%
\bibitem [{\citenamefont {Bai}\ \emph {et~al.}(2020)\citenamefont {Bai},
  \citenamefont {Koltun},\ and\ \citenamefont {Kolter}}]{Bai_multiscale}%
  \BibitemOpen
  \bibfield  {author} {\bibinfo {author} {\bibfnamefont {S.}~\bibnamefont
  {Bai}}, \bibinfo {author} {\bibfnamefont {V.}~\bibnamefont {Koltun}}, \ and\
  \bibinfo {author} {\bibfnamefont {J.~Z.}\ \bibnamefont {Kolter}},\ }in\ \href
  {https://proceedings.neurips.cc/paper/2020/file/3812f9a59b634c2a9c574610eaba5bed-Paper.pdf}
  {\emph {\bibinfo {booktitle} {Advances in Neural Information Processing
  Systems}}},\ Vol.~\bibinfo {volume} {33},\ \bibinfo {editor} {edited by\
  \bibinfo {editor} {\bibfnamefont {H.}~\bibnamefont {Larochelle}}, \bibinfo
  {editor} {\bibfnamefont {M.}~\bibnamefont {Ranzato}}, \bibinfo {editor}
  {\bibfnamefont {R.}~\bibnamefont {Hadsell}}, \bibinfo {editor} {\bibfnamefont
  {M.}~\bibnamefont {Balcan}}, \ and\ \bibinfo {editor} {\bibfnamefont
  {H.}~\bibnamefont {Lin}}}\ (\bibinfo  {publisher} {Curran Associates, Inc.},\
  \bibinfo {year} {2020})\ pp.\ \bibinfo {pages} {5238--5250}\BibitemShut
  {NoStop}%
\bibitem [{\citenamefont {Gillespie}(2007)}]{gillespie_stochastic}%
  \BibitemOpen
  \bibfield  {author} {\bibinfo {author} {\bibfnamefont {D.~T.}\ \bibnamefont
  {Gillespie}},\ }\href {\doibase 10.1146/annurev.physchem.58.032806.104637}
  {\bibfield  {journal} {\bibinfo  {journal} {Annual Review of Physical
  Chemistry}\ }\textbf {\bibinfo {volume} {58}},\ \bibinfo {pages} {35}
  (\bibinfo {year} {2007})}\BibitemShut {NoStop}%
\bibitem [{\citenamefont {Spall}(1998{\natexlab{a}})}]{spall_overview_1998}%
  \BibitemOpen
  \bibfield  {author} {\bibinfo {author} {\bibfnamefont {J.~C.}\ \bibnamefont
  {Spall}},\ }\href@noop {} {\bibfield  {journal} {\bibinfo  {journal} {Johns
  Hopkins apl technical digest}\ }\textbf {\bibinfo {volume} {19}},\ \bibinfo
  {pages} {482} (\bibinfo {year} {1998}{\natexlab{a}})}\BibitemShut {NoStop}%
\bibitem [{\citenamefont {Greydanus}(2020)}]{greydanus2020scaling}%
  \BibitemOpen
  \bibfield  {author} {\bibinfo {author} {\bibfnamefont {S.}~\bibnamefont
  {Greydanus}},\ }\href@noop {} {\bibfield  {journal} {\bibinfo  {journal}
  {arXiv preprint arXiv:2011.14439}\ } (\bibinfo {year} {2020})}\BibitemShut
  {NoStop}%
\bibitem [{\citenamefont {Gardner}\ and\ \citenamefont
  {Derrida}(1988)}]{Gardner_optimal}%
  \BibitemOpen
  \bibfield  {author} {\bibinfo {author} {\bibfnamefont {E.}~\bibnamefont
  {Gardner}}\ and\ \bibinfo {author} {\bibfnamefont {B.}~\bibnamefont
  {Derrida}},\ }\href {\doibase 10.1088/0305-4470/21/1/031} {\bibfield
  {journal} {\bibinfo  {journal} {Journal of Physics A: Mathematical and
  General}\ }\textbf {\bibinfo {volume} {21}},\ \bibinfo {pages} {271}
  (\bibinfo {year} {1988})}\BibitemShut {NoStop}%
\bibitem [{\citenamefont {Gardner}\ and\ \citenamefont
  {Derrida}(1989)}]{Gardner_unfinished}%
  \BibitemOpen
  \bibfield  {author} {\bibinfo {author} {\bibfnamefont {E.}~\bibnamefont
  {Gardner}}\ and\ \bibinfo {author} {\bibfnamefont {B.}~\bibnamefont
  {Derrida}},\ }\href {\doibase 10.1088/0305-4470/22/12/004} {\bibfield
  {journal} {\bibinfo  {journal} {Journal of Physics A: Mathematical and
  General}\ }\textbf {\bibinfo {volume} {22}},\ \bibinfo {pages} {1983}
  (\bibinfo {year} {1989})}\BibitemShut {NoStop}%
\bibitem [{\citenamefont {Engel}\ and\ \citenamefont {Van~den
  Broeck}(2001{\natexlab{a}})}]{engel_van_den_broeck_2001}%
  \BibitemOpen
  \bibfield  {author} {\bibinfo {author} {\bibfnamefont {A.}~\bibnamefont
  {Engel}}\ and\ \bibinfo {author} {\bibfnamefont {C.}~\bibnamefont {Van~den
  Broeck}},\ }\href {\doibase 10.1017/CBO9781139164542} {\emph {\bibinfo
  {title} {Statistical Mechanics of Learning}}}\ (\bibinfo  {publisher}
  {Cambridge University Press},\ \bibinfo {year} {2001})\BibitemShut {NoStop}%
\bibitem [{\citenamefont {Schwarze}\ and\ \citenamefont
  {Hertz}(1992)}]{schwarze1992generalization}%
  \BibitemOpen
  \bibfield  {author} {\bibinfo {author} {\bibfnamefont {H.}~\bibnamefont
  {Schwarze}}\ and\ \bibinfo {author} {\bibfnamefont {J.}~\bibnamefont
  {Hertz}},\ }\href@noop {} {\bibfield  {journal} {\bibinfo  {journal} {EPL
  (Europhysics Letters)}\ }\textbf {\bibinfo {volume} {20}},\ \bibinfo {pages}
  {375} (\bibinfo {year} {1992})}\BibitemShut {NoStop}%
\bibitem [{\citenamefont {Seung}\ \emph {et~al.}(1992)\citenamefont {Seung},
  \citenamefont {Sompolinsky},\ and\ \citenamefont
  {Tishby}}]{stat_mech_seung_sompo}%
  \BibitemOpen
  \bibfield  {author} {\bibinfo {author} {\bibfnamefont {H.~S.}\ \bibnamefont
  {Seung}}, \bibinfo {author} {\bibfnamefont {H.}~\bibnamefont {Sompolinsky}},
  \ and\ \bibinfo {author} {\bibfnamefont {N.}~\bibnamefont {Tishby}},\ }\href
  {\doibase 10.1103/PhysRevA.45.6056} {\bibfield  {journal} {\bibinfo
  {journal} {Phys. Rev. A}\ }\textbf {\bibinfo {volume} {45}},\ \bibinfo
  {pages} {6056} (\bibinfo {year} {1992})}\BibitemShut {NoStop}%
\bibitem [{\citenamefont {Engel}\ and\ \citenamefont {Van~den
  Broeck}(2001{\natexlab{b}})}]{engel2001statistical}%
  \BibitemOpen
  \bibfield  {author} {\bibinfo {author} {\bibfnamefont {A.}~\bibnamefont
  {Engel}}\ and\ \bibinfo {author} {\bibfnamefont {C.}~\bibnamefont {Van~den
  Broeck}},\ }\href@noop {} {\emph {\bibinfo {title} {Statistical mechanics of
  learning}}}\ (\bibinfo  {publisher} {Cambridge University Press},\ \bibinfo
  {year} {2001})\BibitemShut {NoStop}%
\bibitem [{\citenamefont {Baldassi}\ \emph {et~al.}(2015)\citenamefont
  {Baldassi}, \citenamefont {Ingrosso}, \citenamefont {Lucibello},
  \citenamefont {Saglietti},\ and\ \citenamefont
  {Zecchina}}]{baldassi_subdominant_2015}%
  \BibitemOpen
  \bibfield  {author} {\bibinfo {author} {\bibfnamefont {C.}~\bibnamefont
  {Baldassi}}, \bibinfo {author} {\bibfnamefont {A.}~\bibnamefont {Ingrosso}},
  \bibinfo {author} {\bibfnamefont {C.}~\bibnamefont {Lucibello}}, \bibinfo
  {author} {\bibfnamefont {L.}~\bibnamefont {Saglietti}}, \ and\ \bibinfo
  {author} {\bibfnamefont {R.}~\bibnamefont {Zecchina}},\ }\href {\doibase
  10.1103/PhysRevLett.115.128101} {\bibfield  {journal} {\bibinfo  {journal}
  {Phys. Rev. Lett.}\ }\textbf {\bibinfo {volume} {115}},\ \bibinfo {pages}
  {128101} (\bibinfo {year} {2015})}\BibitemShut {NoStop}%
\bibitem [{\citenamefont {Loureiro}\ \emph {et~al.}(2021)\citenamefont
  {Loureiro}, \citenamefont {Sicuro}, \citenamefont {Gerbelot}, \citenamefont
  {Pacco}, \citenamefont {Krzakala},\ and\ \citenamefont
  {Zdeborov\'{a}}}]{Louriero_mixture}%
  \BibitemOpen
  \bibfield  {author} {\bibinfo {author} {\bibfnamefont {B.}~\bibnamefont
  {Loureiro}}, \bibinfo {author} {\bibfnamefont {G.}~\bibnamefont {Sicuro}},
  \bibinfo {author} {\bibfnamefont {C.}~\bibnamefont {Gerbelot}}, \bibinfo
  {author} {\bibfnamefont {A.}~\bibnamefont {Pacco}}, \bibinfo {author}
  {\bibfnamefont {F.}~\bibnamefont {Krzakala}}, \ and\ \bibinfo {author}
  {\bibfnamefont {L.}~\bibnamefont {Zdeborov\'{a}}},\ }in\ \href
  {https://proceedings.neurips.cc/paper/2021/file/543e83748234f7cbab21aa0ade66565f-Paper.pdf}
  {\emph {\bibinfo {booktitle} {Advances in Neural Information Processing
  Systems}}},\ Vol.~\bibinfo {volume} {34},\ \bibinfo {editor} {edited by\
  \bibinfo {editor} {\bibfnamefont {M.}~\bibnamefont {Ranzato}}, \bibinfo
  {editor} {\bibfnamefont {A.}~\bibnamefont {Beygelzimer}}, \bibinfo {editor}
  {\bibfnamefont {Y.}~\bibnamefont {Dauphin}}, \bibinfo {editor} {\bibfnamefont
  {P.}~\bibnamefont {Liang}}, \ and\ \bibinfo {editor} {\bibfnamefont {J.~W.}\
  \bibnamefont {Vaughan}}}\ (\bibinfo  {publisher} {Curran Associates, Inc.},\
  \bibinfo {year} {2021})\ pp.\ \bibinfo {pages} {10144--10157}\BibitemShut
  {NoStop}%
\bibitem [{\citenamefont {Refinetti}\ \emph {et~al.}(2021)\citenamefont
  {Refinetti}, \citenamefont {Goldt}, \citenamefont {Krzakala},\ and\
  \citenamefont {Zdeborova}}]{refinetti2021classifying}%
  \BibitemOpen
  \bibfield  {author} {\bibinfo {author} {\bibfnamefont {M.}~\bibnamefont
  {Refinetti}}, \bibinfo {author} {\bibfnamefont {S.}~\bibnamefont {Goldt}},
  \bibinfo {author} {\bibfnamefont {F.}~\bibnamefont {Krzakala}}, \ and\
  \bibinfo {author} {\bibfnamefont {L.}~\bibnamefont {Zdeborova}},\ }in\ \href
  {http://proceedings.mlr.press/v139/refinetti21b.html} {\emph {\bibinfo
  {booktitle} {Proceedings of the 38th International Conference on Machine
  Learning}}},\ \bibinfo {series} {Proceedings of Machine Learning Research},
  Vol.\ \bibinfo {volume} {139},\ \bibinfo {editor} {edited by\ \bibinfo
  {editor} {\bibfnamefont {M.}~\bibnamefont {Meila}}\ and\ \bibinfo {editor}
  {\bibfnamefont {T.}~\bibnamefont {Zhang}}}\ (\bibinfo  {publisher} {PMLR},\
  \bibinfo {year} {2021})\ pp.\ \bibinfo {pages} {8936--8947}\BibitemShut
  {NoStop}%
\bibitem [{\citenamefont {Tishby}\ \emph {et~al.}(1999)\citenamefont {Tishby},
  \citenamefont {Pereira},\ and\ \citenamefont
  {Bialek}}]{tishby_information_2000}%
  \BibitemOpen
  \bibfield  {author} {\bibinfo {author} {\bibfnamefont {N.}~\bibnamefont
  {Tishby}}, \bibinfo {author} {\bibfnamefont {F.~C.}\ \bibnamefont {Pereira}},
  \ and\ \bibinfo {author} {\bibfnamefont {W.}~\bibnamefont {Bialek}},\ }in\
  \href {https://arxiv.org/abs/physics/0004057} {\emph {\bibinfo {booktitle}
  {Proc. of the 37-th Annual Allerton Conference on Communication, Control and
  Computing}}}\ (\bibinfo {year} {1999})\ pp.\ \bibinfo {pages}
  {368--377}\BibitemShut {NoStop}%
\bibitem [{\citenamefont {Strouse}\ and\ \citenamefont
  {Schwab}(2017)}]{strouse_deterministic_2017}%
  \BibitemOpen
  \bibfield  {author} {\bibinfo {author} {\bibfnamefont {D.}~\bibnamefont
  {Strouse}}\ and\ \bibinfo {author} {\bibfnamefont {D.~J.}\ \bibnamefont
  {Schwab}},\ }\href {\doibase 10.1162/NECO_a_00961} {\bibfield  {journal}
  {\bibinfo  {journal} {Neural Computation}\ }\textbf {\bibinfo {volume}
  {29}},\ \bibinfo {pages} {1611} (\bibinfo {year} {2017})}\BibitemShut
  {NoStop}%
\bibitem [{\citenamefont {Chalk}\ \emph {et~al.}(2016)\citenamefont {Chalk},
  \citenamefont {Marre},\ and\ \citenamefont {Tkacik}}]{chalk_relevant_2016}%
  \BibitemOpen
  \bibfield  {author} {\bibinfo {author} {\bibfnamefont {M.}~\bibnamefont
  {Chalk}}, \bibinfo {author} {\bibfnamefont {O.}~\bibnamefont {Marre}}, \ and\
  \bibinfo {author} {\bibfnamefont {G.}~\bibnamefont {Tkacik}},\ }in\ \href
  {https://proceedings.neurips.cc/paper/2016/file/a89cf525e1d9f04d16ce31165e139a4b-Paper.pdf}
  {\emph {\bibinfo {booktitle} {Advances in Neural Information Processing
  Systems}}},\ Vol.~\bibinfo {volume} {29},\ \bibinfo {editor} {edited by\
  \bibinfo {editor} {\bibfnamefont {D.}~\bibnamefont {Lee}}, \bibinfo {editor}
  {\bibfnamefont {M.}~\bibnamefont {Sugiyama}}, \bibinfo {editor}
  {\bibfnamefont {U.}~\bibnamefont {Luxburg}}, \bibinfo {editor} {\bibfnamefont
  {I.}~\bibnamefont {Guyon}}, \ and\ \bibinfo {editor} {\bibfnamefont
  {R.}~\bibnamefont {Garnett}}}\ (\bibinfo  {publisher} {Curran Associates,
  Inc.},\ \bibinfo {year} {2016})\BibitemShut {NoStop}%
\bibitem [{\citenamefont {Baiesi}\ and\ \citenamefont
  {Maes}(2018)}]{baiesi2018life}%
  \BibitemOpen
  \bibfield  {author} {\bibinfo {author} {\bibfnamefont {M.}~\bibnamefont
  {Baiesi}}\ and\ \bibinfo {author} {\bibfnamefont {C.}~\bibnamefont {Maes}},\
  }\href@noop {} {\bibfield  {journal} {\bibinfo  {journal} {Journal of Physics
  Communications}\ }\textbf {\bibinfo {volume} {2}},\ \bibinfo {pages} {045017}
  (\bibinfo {year} {2018})}\BibitemShut {NoStop}%
\bibitem [{\citenamefont {Hatano}\ and\ \citenamefont
  {Sasa}(2001)}]{hatano_steady-state_2001}%
  \BibitemOpen
  \bibfield  {author} {\bibinfo {author} {\bibfnamefont {T.}~\bibnamefont
  {Hatano}}\ and\ \bibinfo {author} {\bibfnamefont {S.-i.}\ \bibnamefont
  {Sasa}},\ }\href {\doibase 10.1103/PhysRevLett.86.3463} {\bibfield  {journal}
  {\bibinfo  {journal} {Physical Review Letters}\ }\textbf {\bibinfo {volume}
  {86}},\ \bibinfo {pages} {3463} (\bibinfo {year} {2001})}\BibitemShut
  {NoStop}%
\bibitem [{\citenamefont {Vaswani}\ \emph
  {et~al.}(2017{\natexlab{a}})\citenamefont {Vaswani}, \citenamefont {Shazeer},
  \citenamefont {Parmar}, \citenamefont {Uszkoreit}, \citenamefont {Jones},
  \citenamefont {Gomez}, \citenamefont {Kaiser},\ and\ \citenamefont
  {Polosukhin}}]{vaswani2017attention}%
  \BibitemOpen
  \bibfield  {author} {\bibinfo {author} {\bibfnamefont {A.}~\bibnamefont
  {Vaswani}}, \bibinfo {author} {\bibfnamefont {N.}~\bibnamefont {Shazeer}},
  \bibinfo {author} {\bibfnamefont {N.}~\bibnamefont {Parmar}}, \bibinfo
  {author} {\bibfnamefont {J.}~\bibnamefont {Uszkoreit}}, \bibinfo {author}
  {\bibfnamefont {L.}~\bibnamefont {Jones}}, \bibinfo {author} {\bibfnamefont
  {A.~N.}\ \bibnamefont {Gomez}}, \bibinfo {author} {\bibfnamefont
  {{\L}.}~\bibnamefont {Kaiser}}, \ and\ \bibinfo {author} {\bibfnamefont
  {I.}~\bibnamefont {Polosukhin}},\ }\href@noop {} {\bibfield  {journal}
  {\bibinfo  {journal} {Advances in neural information processing systems}\
  }\textbf {\bibinfo {volume} {30}} (\bibinfo {year}
  {2017}{\natexlab{a}})}\BibitemShut {NoStop}%
\bibitem [{\citenamefont {Devlin}\ \emph {et~al.}(2018)\citenamefont {Devlin},
  \citenamefont {Chang}, \citenamefont {Lee},\ and\ \citenamefont
  {Toutanova}}]{devlin2018bert}%
  \BibitemOpen
  \bibfield  {author} {\bibinfo {author} {\bibfnamefont {J.}~\bibnamefont
  {Devlin}}, \bibinfo {author} {\bibfnamefont {M.-W.}\ \bibnamefont {Chang}},
  \bibinfo {author} {\bibfnamefont {K.}~\bibnamefont {Lee}}, \ and\ \bibinfo
  {author} {\bibfnamefont {K.}~\bibnamefont {Toutanova}},\ }\href@noop {}
  {\bibfield  {journal} {\bibinfo  {journal} {arXiv preprint arXiv:1810.04805}\
  } (\bibinfo {year} {2018})}\BibitemShut {NoStop}%
\bibitem [{\citenamefont {OpenAI}(2023)}]{openai2023gpt4}%
  \BibitemOpen
  \bibfield  {author} {\bibinfo {author} {\bibnamefont {OpenAI}},\ }\href@noop
  {} {\enquote {\bibinfo {title} {Gpt-4 technical report},}\ } (\bibinfo {year}
  {2023}),\ \Eprint {http://arxiv.org/abs/2303.08774} {arXiv:2303.08774
  [cs.CL]} \BibitemShut {NoStop}%
\bibitem [{\citenamefont {Hopfield}(1982)}]{hopfield_neural_1982}%
  \BibitemOpen
  \bibfield  {author} {\bibinfo {author} {\bibfnamefont {J.~J.}\ \bibnamefont
  {Hopfield}},\ }\href {\doibase 10.1073/pnas.79.8.2554} {\bibfield  {journal}
  {\bibinfo  {journal} {Proceedings of the National Academy of Sciences}\
  }\textbf {\bibinfo {volume} {79}},\ \bibinfo {pages} {2554} (\bibinfo {year}
  {1982})}\BibitemShut {NoStop}%
\bibitem [{\citenamefont {Krotov}\ and\ \citenamefont
  {Hopfield}(2016)}]{krotov_dense}%
  \BibitemOpen
  \bibfield  {author} {\bibinfo {author} {\bibfnamefont {D.}~\bibnamefont
  {Krotov}}\ and\ \bibinfo {author} {\bibfnamefont {J.~J.}\ \bibnamefont
  {Hopfield}},\ }in\ \href
  {https://proceedings.neurips.cc/paper_files/paper/2016/file/eaae339c4d89fc102edd9dbdb6a28915-Paper.pdf}
  {\emph {\bibinfo {booktitle} {Advances in Neural Information Processing
  Systems}}},\ Vol.~\bibinfo {volume} {29},\ \bibinfo {editor} {edited by\
  \bibinfo {editor} {\bibfnamefont {D.}~\bibnamefont {Lee}}, \bibinfo {editor}
  {\bibfnamefont {M.}~\bibnamefont {Sugiyama}}, \bibinfo {editor}
  {\bibfnamefont {U.}~\bibnamefont {Luxburg}}, \bibinfo {editor} {\bibfnamefont
  {I.}~\bibnamefont {Guyon}}, \ and\ \bibinfo {editor} {\bibfnamefont
  {R.}~\bibnamefont {Garnett}}}\ (\bibinfo  {publisher} {Curran Associates,
  Inc.},\ \bibinfo {year} {2016})\BibitemShut {NoStop}%
\bibitem [{\citenamefont {Krotov}\ and\ \citenamefont
  {Hopfield}(2021)}]{krotov2021large}%
  \BibitemOpen
  \bibfield  {author} {\bibinfo {author} {\bibfnamefont {D.}~\bibnamefont
  {Krotov}}\ and\ \bibinfo {author} {\bibfnamefont {J.}~\bibnamefont
  {Hopfield}},\ }\href@noop {} {\enquote {\bibinfo {title} {Large associative
  memory problem in neurobiology and machine learning},}\ } (\bibinfo {year}
  {2021}),\ \Eprint {http://arxiv.org/abs/2008.06996} {arXiv:2008.06996
  [q-bio.NC]} \BibitemShut {NoStop}%
\bibitem [{\citenamefont {Ramsauer}\ \emph {et~al.}(2021)\citenamefont
  {Ramsauer}, \citenamefont {Schäfl}, \citenamefont {Lehner}, \citenamefont
  {Seidl}, \citenamefont {Widrich}, \citenamefont {Adler}, \citenamefont
  {Gruber}, \citenamefont {Holzleitner}, \citenamefont {Pavlović},
  \citenamefont {Sandve}, \citenamefont {Greiff}, \citenamefont {Kreil},
  \citenamefont {Kopp}, \citenamefont {Klambauer}, \citenamefont
  {Brandstetter},\ and\ \citenamefont {Hochreiter}}]{ramsauer2021hopfield}%
  \BibitemOpen
  \bibfield  {author} {\bibinfo {author} {\bibfnamefont {H.}~\bibnamefont
  {Ramsauer}}, \bibinfo {author} {\bibfnamefont {B.}~\bibnamefont {Schäfl}},
  \bibinfo {author} {\bibfnamefont {J.}~\bibnamefont {Lehner}}, \bibinfo
  {author} {\bibfnamefont {P.}~\bibnamefont {Seidl}}, \bibinfo {author}
  {\bibfnamefont {M.}~\bibnamefont {Widrich}}, \bibinfo {author} {\bibfnamefont
  {T.}~\bibnamefont {Adler}}, \bibinfo {author} {\bibfnamefont
  {L.}~\bibnamefont {Gruber}}, \bibinfo {author} {\bibfnamefont
  {M.}~\bibnamefont {Holzleitner}}, \bibinfo {author} {\bibfnamefont
  {M.}~\bibnamefont {Pavlović}}, \bibinfo {author} {\bibfnamefont {G.~K.}\
  \bibnamefont {Sandve}}, \bibinfo {author} {\bibfnamefont {V.}~\bibnamefont
  {Greiff}}, \bibinfo {author} {\bibfnamefont {D.}~\bibnamefont {Kreil}},
  \bibinfo {author} {\bibfnamefont {M.}~\bibnamefont {Kopp}}, \bibinfo {author}
  {\bibfnamefont {G.}~\bibnamefont {Klambauer}}, \bibinfo {author}
  {\bibfnamefont {J.}~\bibnamefont {Brandstetter}}, \ and\ \bibinfo {author}
  {\bibfnamefont {S.}~\bibnamefont {Hochreiter}},\ }\href@noop {} {\enquote
  {\bibinfo {title} {Hopfield networks is all you need},}\ } (\bibinfo {year}
  {2021}),\ \Eprint {http://arxiv.org/abs/2008.02217} {arXiv:2008.02217
  [cs.NE]} \BibitemShut {NoStop}%
\bibitem [{\citenamefont {Vaswani}\ \emph
  {et~al.}(2017{\natexlab{b}})\citenamefont {Vaswani}, \citenamefont {Shazeer},
  \citenamefont {Parmar}, \citenamefont {Uszkoreit}, \citenamefont {Jones},
  \citenamefont {Gomez}, \citenamefont {Kaiser},\ and\ \citenamefont
  {Polosukhin}}]{attention_allyouneed}%
  \BibitemOpen
  \bibfield  {author} {\bibinfo {author} {\bibfnamefont {A.}~\bibnamefont
  {Vaswani}}, \bibinfo {author} {\bibfnamefont {N.}~\bibnamefont {Shazeer}},
  \bibinfo {author} {\bibfnamefont {N.}~\bibnamefont {Parmar}}, \bibinfo
  {author} {\bibfnamefont {J.}~\bibnamefont {Uszkoreit}}, \bibinfo {author}
  {\bibfnamefont {L.}~\bibnamefont {Jones}}, \bibinfo {author} {\bibfnamefont
  {A.~N.}\ \bibnamefont {Gomez}}, \bibinfo {author} {\bibfnamefont {L.~u.}\
  \bibnamefont {Kaiser}}, \ and\ \bibinfo {author} {\bibfnamefont
  {I.}~\bibnamefont {Polosukhin}},\ }in\ \href
  {https://proceedings.neurips.cc/paper_files/paper/2017/file/3f5ee243547dee91fbd053c1c4a845aa-Paper.pdf}
  {\emph {\bibinfo {booktitle} {Advances in Neural Information Processing
  Systems}}},\ Vol.~\bibinfo {volume} {30},\ \bibinfo {editor} {edited by\
  \bibinfo {editor} {\bibfnamefont {I.}~\bibnamefont {Guyon}}, \bibinfo
  {editor} {\bibfnamefont {U.~V.}\ \bibnamefont {Luxburg}}, \bibinfo {editor}
  {\bibfnamefont {S.}~\bibnamefont {Bengio}}, \bibinfo {editor} {\bibfnamefont
  {H.}~\bibnamefont {Wallach}}, \bibinfo {editor} {\bibfnamefont
  {R.}~\bibnamefont {Fergus}}, \bibinfo {editor} {\bibfnamefont
  {S.}~\bibnamefont {Vishwanathan}}, \ and\ \bibinfo {editor} {\bibfnamefont
  {R.}~\bibnamefont {Garnett}}}\ (\bibinfo  {publisher} {Curran Associates,
  Inc.},\ \bibinfo {year} {2017})\BibitemShut {NoStop}%
\bibitem [{\citenamefont {Russo}\ and\ \citenamefont
  {Roy}(2016)}]{russo_information-theoretic_nodate}%
  \BibitemOpen
  \bibfield  {author} {\bibinfo {author} {\bibfnamefont {D.}~\bibnamefont
  {Russo}}\ and\ \bibinfo {author} {\bibfnamefont {B.~V.}\ \bibnamefont
  {Roy}},\ }\href {http://jmlr.org/papers/v17/14-087.html} {\bibfield
  {journal} {\bibinfo  {journal} {Journal of Machine Learning Research}\
  }\textbf {\bibinfo {volume} {17}},\ \bibinfo {pages} {1} (\bibinfo {year}
  {2016})}\BibitemShut {NoStop}%
\bibitem [{\citenamefont {Xu}\ and\ \citenamefont
  {Raginsky}(2017)}]{xu_information-theoretic_2017}%
  \BibitemOpen
  \bibfield  {author} {\bibinfo {author} {\bibfnamefont {A.}~\bibnamefont
  {Xu}}\ and\ \bibinfo {author} {\bibfnamefont {M.}~\bibnamefont {Raginsky}},\
  }in\ \href
  {https://proceedings.neurips.cc/paper/2017/file/ad71c82b22f4f65b9398f76d8be4c615-Paper.pdf}
  {\emph {\bibinfo {booktitle} {Advances in Neural Information Processing
  Systems}}},\ Vol.~\bibinfo {volume} {30},\ \bibinfo {editor} {edited by\
  \bibinfo {editor} {\bibfnamefont {I.}~\bibnamefont {Guyon}}, \bibinfo
  {editor} {\bibfnamefont {U.~V.}\ \bibnamefont {Luxburg}}, \bibinfo {editor}
  {\bibfnamefont {S.}~\bibnamefont {Bengio}}, \bibinfo {editor} {\bibfnamefont
  {H.}~\bibnamefont {Wallach}}, \bibinfo {editor} {\bibfnamefont
  {R.}~\bibnamefont {Fergus}}, \bibinfo {editor} {\bibfnamefont
  {S.}~\bibnamefont {Vishwanathan}}, \ and\ \bibinfo {editor} {\bibfnamefont
  {R.}~\bibnamefont {Garnett}}}\ (\bibinfo  {publisher} {Curran Associates,
  Inc.},\ \bibinfo {year} {2017})\BibitemShut {NoStop}%
\bibitem [{\citenamefont {Russo}\ and\ \citenamefont
  {Van~Roy}(0)}]{russo_satisficing_2020}%
  \BibitemOpen
  \bibfield  {author} {\bibinfo {author} {\bibfnamefont {D.}~\bibnamefont
  {Russo}}\ and\ \bibinfo {author} {\bibfnamefont {B.}~\bibnamefont
  {Van~Roy}},\ }\href {\doibase 10.1287/moor.2021.1229} {\bibfield  {journal}
  {\bibinfo  {journal} {Mathematics of Operations Research}\ }\textbf {\bibinfo
  {volume} {0}},\ \bibinfo {pages} {null} (\bibinfo {year} {0})},\ \Eprint
  {http://arxiv.org/abs/https://doi.org/10.1287/moor.2021.1229}
  {https://doi.org/10.1287/moor.2021.1229} \BibitemShut {NoStop}%
\bibitem [{\citenamefont {Pacelli}\ and\ \citenamefont
  {Majumdar}(2019)}]{pacelli_task-driven_2019}%
  \BibitemOpen
  \bibfield  {author} {\bibinfo {author} {\bibfnamefont {V.}~\bibnamefont
  {Pacelli}}\ and\ \bibinfo {author} {\bibfnamefont {A.}~\bibnamefont
  {Majumdar}},\ }in\ \href {\doibase 10.1109/ICRA.2019.8794213} {\emph
  {\bibinfo {booktitle} {2019 International Conference on Robotics and
  Automation (ICRA)}}}\ (\bibinfo {year} {2019})\ pp.\ \bibinfo {pages}
  {2061--2067}\BibitemShut {NoStop}%
\bibitem [{\citenamefont {Pacelli}\ and\ \citenamefont
  {Majumdar}(2020)}]{pacelli_learning_2020}%
  \BibitemOpen
  \bibfield  {author} {\bibinfo {author} {\bibfnamefont {V.}~\bibnamefont
  {Pacelli}}\ and\ \bibinfo {author} {\bibfnamefont {A.}~\bibnamefont
  {Majumdar}},\ }in\ \href {\doibase 10.15607/RSS.2020.XVI.101} {\emph
  {\bibinfo {booktitle} {Robotics: {Science} and {Systems} {XVI}}}}\ (\bibinfo
  {publisher} {Robotics: Science and Systems Foundation},\ \bibinfo {year}
  {2020})\BibitemShut {NoStop}%
\bibitem [{\citenamefont {Majumdar}\ and\ \citenamefont
  {Pacelli}(2022)}]{majumdar_fundamental_2022}%
  \BibitemOpen
  \bibfield  {author} {\bibinfo {author} {\bibfnamefont {A.}~\bibnamefont
  {Majumdar}}\ and\ \bibinfo {author} {\bibfnamefont {V.}~\bibnamefont
  {Pacelli}},\ }\href {http://arxiv.org/abs/2202.00129} {\enquote {\bibinfo
  {title} {Fundamental {Performance} {Limits} for {Sensor}-{Based} {Robot}
  {Control} and {Policy} {Learning}},}\ } (\bibinfo {year} {2022}),\ \bibinfo
  {note} {arXiv:2202.00129 [cs, math]}\BibitemShut {NoStop}%
\bibitem [{\citenamefont {Gillespie}(1976)}]{gillespie_general}%
  \BibitemOpen
  \bibfield  {author} {\bibinfo {author} {\bibfnamefont {D.~T.}\ \bibnamefont
  {Gillespie}},\ }\href {\doibase https://doi.org/10.1016/0021-9991(76)90041-3}
  {\bibfield  {journal} {\bibinfo  {journal} {Journal of Computational
  Physics}\ }\textbf {\bibinfo {volume} {22}},\ \bibinfo {pages} {403}
  (\bibinfo {year} {1976})}\BibitemShut {NoStop}%
\bibitem [{\citenamefont {Martynec}\ \emph {et~al.}(2020)\citenamefont
  {Martynec}, \citenamefont {Klapp},\ and\ \citenamefont {Loos}}]{Loos_potts}%
  \BibitemOpen
  \bibfield  {author} {\bibinfo {author} {\bibfnamefont {T.}~\bibnamefont
  {Martynec}}, \bibinfo {author} {\bibfnamefont {S.~H.~L.}\ \bibnamefont
  {Klapp}}, \ and\ \bibinfo {author} {\bibfnamefont {S.~A.~M.}\ \bibnamefont
  {Loos}},\ }\href {\doibase 10.1088/1367-2630/abb5f0} {\bibfield  {journal}
  {\bibinfo  {journal} {New Journal of Physics}\ }\textbf {\bibinfo {volume}
  {22}},\ \bibinfo {pages} {093069} (\bibinfo {year} {2020})}\BibitemShut
  {NoStop}%
\bibitem [{\citenamefont {Spall}(1992)}]{spsa_multivariate}%
  \BibitemOpen
  \bibfield  {author} {\bibinfo {author} {\bibfnamefont {J.}~\bibnamefont
  {Spall}},\ }\href {\doibase 10.1109/9.119632} {\bibfield  {journal} {\bibinfo
   {journal} {IEEE Transactions on Automatic Control}\ }\textbf {\bibinfo
  {volume} {37}},\ \bibinfo {pages} {332} (\bibinfo {year} {1992})}\BibitemShut
  {NoStop}%
\bibitem [{\citenamefont {Spall}(1998{\natexlab{b}})}]{spsa_implementation}%
  \BibitemOpen
  \bibfield  {author} {\bibinfo {author} {\bibfnamefont {J.}~\bibnamefont
  {Spall}},\ }\href {\doibase 10.1109/7.705889} {\bibfield  {journal} {\bibinfo
   {journal} {IEEE Transactions on Aerospace and Electronic Systems}\ }\textbf
  {\bibinfo {volume} {34}},\ \bibinfo {pages} {817} (\bibinfo {year}
  {1998}{\natexlab{b}})}\BibitemShut {NoStop}%
\bibitem [{\citenamefont {Seara}\ \emph {et~al.}(2021)\citenamefont {Seara},
  \citenamefont {Machta},\ and\ \citenamefont
  {Murrell}}]{seara_irreversibility_2021}%
  \BibitemOpen
  \bibfield  {author} {\bibinfo {author} {\bibfnamefont {D.~S.}\ \bibnamefont
  {Seara}}, \bibinfo {author} {\bibfnamefont {B.~B.}\ \bibnamefont {Machta}}, \
  and\ \bibinfo {author} {\bibfnamefont {M.~P.}\ \bibnamefont {Murrell}},\
  }\href {\doibase 10.1038/s41467-020-20281-2} {\bibfield  {journal} {\bibinfo
  {journal} {Nature Communications}\ }\textbf {\bibinfo {volume} {12}},\
  \bibinfo {pages} {392} (\bibinfo {year} {2021})}\BibitemShut {NoStop}%
\end{thebibliography}%

\end{document}